\documentclass[twocolumn,letterpaper]{IEEEAerospaceCLS}  

\usepackage{graphicx}
\usepackage{booktabs} 
\usepackage[cmex10]{amsmath}
\usepackage{epstopdf}
\usepackage[short]{optidef}
\usepackage[utf8]{inputenc}
\usepackage[english]{babel}
\usepackage{color}
\usepackage{epstopdf}
\usepackage{commath}
\usepackage{algorithm}
\usepackage{lipsum}
\usepackage{algpseudocode}
\usepackage{mathptmx}
\usepackage{bm}
\usepackage[caption=false]{subfig}
\usepackage{cite}
\usepackage{gensymb}
\usepackage{amssymb}
\usepackage{mathtools}  
\usepackage{amsfonts}
\usepackage{eucal}
\usepackage{enumitem}
\DeclarePairedDelimiter\floor{\lfloor}{\rfloor}

\begin{document}
\title{Effects of 3D Antenna Radiation and Two-Hop Relaying on Optimal UAV Trajectory in Cellular Networks}

\author{%
Md Moin Uddin Chowdhury, Sung Joon Maeng, Ismail Guvenc\\ 
Department of Electrical and Computer Engineering\\
North Carolina State University\\
Raleigh, NC, 27606\\
\{mchowdh,smaeng,iguvenc\}@ncsu.edu
\and 
Eyuphan Bulut\\ 
Department of Computer Science\\
Virginia Commonwealth University\\
Richmond, VA, 23284\\
ebulut@vcu.edu
\thanks{\footnotesize 978-1-5386-6854-2/19/$\$31.00$ \copyright2019 IEEE}              
}

\maketitle

\thispagestyle{plain}
\pagestyle{plain}

\maketitle

\thispagestyle{plain}
\pagestyle{plain}

\begin{abstract}

In this paper, considering an interference limited in-band downlink cellular network, we study the effects of scheduling criteria,  mobility constraints, path loss models, backhaul constraints, and 3D antenna radiation pattern on trajectory optimization problem of an unmanned aerial vehicle (UAV). In particular, we consider a UAV that is tasked to travel between two locations within a given amount of time (e.g., for delivery or surveillance purposes), and we consider that such a UAV can be used to improve cellular connectivity of mobile users by serving as a relay for the terrestrial network. As the optimization problem is hard to solve numerically, we explore the dynamic programming (DP) technique for finding the optimum UAV trajectory. We utilize capacity and coverage performance of the terrestrial network while studying all the effects of different techniques and phenomenon. Extensive simulations show that the maximum sum-rate trajectory provides the best per user capacity whereas, the optimal proportional fair (PF) rate trajectory provides higher coverage probability than the other two. Since, the generated trajectories are infeasible for the UAV to follow exactly as it can not take sharp turns due to kinematic constraints, we generate smooth trajectory using Bezier curves. Our results show that the cellular capacity using the Bezier curves is close to the capacity observed when using the optimal trajectories.
\end{abstract}

\tableofcontents

\section{Introduction}
Using unmanned aerial vehicles (UAVs) as flying cellular network nodes (aerial base station or aerial user) has gained immense attention from industry and academia in the recent years. Due to their low production cost, flexibility in deployment, and the ability to increase network capacity, UAVs can be deployed as aerial base stations which have been studied in the literature extensively. On the other hand, UAVs can also be used in taking aerial photography after a disaster, package delivery, surveillance, and agriculture. As a matter of fact, while dedicated UAVs can assist cellular service providers to achieve better network performance, the vast number of UAVs that are allotted for other applications can simultaneously assist in providing wireless connectivity to an underlying cellular network. Motivated by this, in this paper, we study the optimal path planning for a UAV acting as a wireless relay for improving the terrestrial downlink cellular network performance. In particular, we consider a UAV that is tasked to travel from one point to another within a given time constraint, and it can also simultaneously provide downlink wireless coverage to terrestrial users  during its mission duration. In order to get the best service from the battery limited UAVs, it is required to optimize the trajectory of the UAVs to provide the best wireless service to users.\\

Considering an underlying cellular network, the optimal UAV path planning has recently been studied extensively in the literature. For instance, authors in \cite{Bulut} find the optimal UAV trajectory by using dynamic programming (DP) \cite{dp}, while maintaining good connectivity with cellular base stations (MBSs). Authors in \cite{adhoc} propose a distributed path planning algorithm for multiple UAVs for delay constrained information delivery to ad-hoc networks. In~\cite{rajeev}, authors consider DP technique to optimize the weighted sum-rate of users in a wireless network. In~\cite{gesbert} authors consider landing spots for recharging to study the trade-off between throughput and battery power using DP. In both works, authors do not consider the presence of other base stations. Authors in \cite{moin}, used DP to find optimal trajectories for different scheduling criterion. In~\cite{rui1}, the minimum throughput of users is maximized in a multi-UAV enabled network, while authors in~\cite{rui2} explore the problem of minimizing the mission completion time  to enable multicasting via trajectory optimization. Energy efficient trajectory optimization using successive convexification techniques is discussed in~\cite{rui3}. Authors in \cite{saad} used deep reinforcement learning to generate trajectory with an aim to reduce interference. A dynamic UAV heading adjustment algorithm for optimizing the ergodic sum-rate of an uplink wireless network was proposed in \cite{jiang} . UAV enabled communication system is also studied in \cite{merwaday}, where optimum UAV locations as well as  interference management parameters are solved. However, none of these studies considered studying the effects of antenna radiation pattern and using the UAVs as relays in interference prevalent cellular networks.

In this paper, we study the effects of scheduling criteria, UAV mobility constraints, antenna radiation pattern, path loss models and backhaul constraints on trajectory optimization problem in an in-band downlink cellular network where, the locations of the ground users (UE) and macro base stations (MBS) follow homogeneous Poisson point processes (PPPs). As the problem is difficult to solve numerically, we use DP to obtain the optimal paths. The generated trajectories by DP are difficult to follow exactly as they involve sharp turns. Hence, we smooth them using Bezier curves. We also study the impact of different path loss models and compare their respective network performances. Moreover, we study the network performances of the trajectories by introducing backhaul constraints, and explore the effect of 3D antenna radiation pattern on the optimal trajectories.

The rest of the paper is organized as follows.~Section 2 illustrates the system model where we discuss our simulation assumptions and DP technique. In Section 3 we discuss the impact of different user association scheduling on optimal trajectories. To generate feasible trajectories we introduce Bezier curve in Section 4 where we study the impact of trajectory smoothening. We discuss three path loss models and study their effects in Section 5. Impacts of backhaul constraint and 3D antenna radiation is discussed in Section 6 and in Section 7, respectively. Finally, Section 8 concludes this paper. 
\section{System Model and Trajectory Optimization Problem}\label{Sec2}

\subsection{Network and UAV Mobility Model}
Let us consider a UAV that is flying at a fixed height $H$ with maximum speed of $V_{\mathrm{max}}$ in a suburban environment. The  UAV has a mission to complete. It has to fly from a start location, $L_{\rm s}$ to a final destination point $L_{\rm f}$ within a fixed time $T$ on an area of $\mathcal{A}$ square meters. Let us consider [$x(s),y(s),h_\textsubscript{uav}$] and [$x(f),y(f),h_\textsubscript{uav}$] to be the 3D Cartesian coordinates of $L_{\rm s}$ and $L_{\rm f}$, respectively. The time-varying horizontal coordinate of the UAV at time instant $t$ is denoted by $r(t)=[x(t),y(t)]^\intercal \in \mathbb{R} \textsuperscript{2x1}$ with $0\leq t\leq T$. The minimum time required for the UAV to reach $L_{\rm f}$ from $L_{\rm s}$ with the maximum speed ${V_{\mathrm{max}}}$ is given by
\begin{equation}
 {T_{\mathrm{min}}} = \frac {\sqrt{(x(s)-x(f))^2+(y(s)-y(f))^2})}{V_{\mathrm{max}}}.    
\end{equation}
The UAV's instantaneous mobility can be modeled as, $\dot{x}(t)=v(t)\cos\phi(t)$ and $\dot{y}(t)=v(t)\sin\phi(t)$,
where $\dot{x}(t)$ and $\dot{y}(t)$ are the time derivative of $x(t)$ and $y(t)$, respectively, $v(t)$ is the velocity, and $\phi(t)$ is the heading angle (in azimuth) of the UAV. Let us also assume that there are $M$ MBSs and $K$ static UEs in the area. The set of the UEs can be denoted as $\mathcal{K}$ with horizontal coordinates  $\textbf{w}_k=[x_k,y_k]^T \in \mathbb{R} \textsuperscript{2x1} , \text{k} \in \mathcal{K}$. The MBS and UE locations follow two identical and independent PPPs. Let us assume the intensities of MBSs and UEs are $\lambda_\textsubscript{mbs}$ and $\lambda_\textsubscript{ue}$ per square km. We also assume that the MBSs transmit with omni-directional antennas and each UE connects to the strongest MBS or the UAV.
\subsection{UAV Trajectory Optimization Problem}
After calculating the SIRs of each UE, the instantaneous logarithmic sum rate of the network at time  $t$ can be expressed as follows:
\begin{equation}
    C(t)= \sum_{k=1}^K \log_{10} {R_{k}(t)},
    \label{eq}
\end{equation}
which is also known as the proportional fair (PF) rate of the network. Here, $R_k(t)$ represents the instantaneous rate of UE $k$ at time $t$. We discuss more about $R_k(t)$ in the \textit{Spectral Efficiency Calculation} subsection.~The PF rate is known to provide a balance between two conflicting performance metrics: throughput and fairness.~Now, we can formulate our trajectory optimization problem over the total mission duration of the UAV as follows:  
\begin{maxi!}	
	  {x(t),y(t)}{\int_{t=0}^{T}C(t)}{}{}\label{Eq3}
	  \addConstraint{\sqrt{\dot{x}(t)^2+\dot{y}(t)^2}}{\leq {V_{\mathrm{max}}}, } \; { t \in [0,T]}
	  \addConstraint{[x(0),y(0)] =}{[x(s),y(s)]}{}
	  \addConstraint{[x(T),y(T)] =}{[x(f),y(f)]}{}.
	  \end{maxi!}
Here, (3b) ensures that the velocity of the UAV does not exceed the maximum limit V\textsubscript{max}, while (3c) and (3d) fix the initial and final location of the mission. Here, we consider $T$ to be as $T \geq T_\textsubscript{min}$, so that there exists at least one feasible solution to the above optimization problem. The maximization problem discussed above is  non-convex problem which is difficult to be solved numerically in general and hence motivated us to use the DP technique to compute the solution.
\subsection{Dynamic Programming Technique}
In this section, the optimization problem in~\eqref{Eq3} is discretized to obtain approximation of the optimal trajectories. The time period $[0,T]$ is divided into $N$ equal intervals of duration $\delta=T/N$ and is indexed by $i=0,....,N-1$. The value of $N$ is chosen so that UAV's position, velocity, and heading angle can be considered constant in an interval. The rate of UE $k$, ${R_{k}(i)}$, at time interval $i$, will be dependent on the distance between the UE and the horizontal position of the UAV. Then, the discrete-time dynamic system can be written as:
\begin{equation}
    \boldsymbol f{r}_{i+1}=\boldsymbol{r}_{i}+ f(i,\boldsymbol{r}_{i},\boldsymbol{u}_{i}),\quad  i=0,1,...,N-1,
\end{equation}
where $\boldsymbol{r}_{i}=[x_i\;y_i]^T$ is the state or the position of the UAV and $\boldsymbol{u}_{i}=[v_i\;\phi_i]^T$ stands for the control action i.e., velocity $v_i$ and the heading angle $\phi_i$, respectively, in the $i$-th time interval. By taking control action at each interval $i$, the UAV will move to next state and it will achieve cost for taking that specific control action. Starting with initial state $r_{0}=[0,0]^T$, the subsequent states can be computed using,
\begin{equation}
    f(i,\boldsymbol{r}_{i},\boldsymbol{u}_{i})= \begin{pmatrix}
  v_i\cos{\phi_i} \\
  v_i\sin{\phi_i}\\
\end{pmatrix}.
\end{equation}
For a given set of control actions $\boldsymbol{\pi} = \{\boldsymbol{u}_{0},\boldsymbol{u}_{1}, . . . ., \boldsymbol{u}_{N-1}\}$,
we can calculate the total cost function starting from state $r_{0}$ as
\begin{equation}
    J_{\pi}(r_0)=J_{r_N}+  \sum_{i=}^{N-1}\sum_{k=1}^K R_{k}(i),
\end{equation}
where,  the terminal cost $J(\boldsymbol{s}_{N})$ is the cost when the UAV reaches the position $[x(f)\; y(f)]$. 

An optimal policy vector ${\pi}^*$ that maximizes the cost is,
\begin{equation}
\label{dp_policy}
 {\pi}^*= \max_{\pi \in \Pi} J_{\pi}(r_0),
\end{equation}	
where $\Pi=\{\boldsymbol{u}_{i},\forall i=0,1,...,N-1 |v_i\leq V_\textsubscript{max},0\leq \phi \leq 360^{\degree} \}$.
 The optimization problem presented above can be solved recursively using Bellman's equations by moving backwards in time as follows~\cite{dp}, \cite{rajeev}:
\begin{equation}
\label{dp_iter}
    J({r}_{i})=\max_{\boldsymbol{u}_{i}}\sum_{k=1}^K \log_{10} {R_{k}(i)}+ J({r}_{i+1}),\quad  i=N-1,..,0.
\end{equation}
The solution of the optimization problem (\ref{dp_policy}), maximizes (\ref{dp_iter}). Since for each state we have to calculate $v_i$ and $\phi_i$, this solution is still computationally expensive.

 

 \subsection{Spectral Efficiency Calculation}
The received power at user $k$ from  MBS $m$  at time $t$, can be calculated as ${S_{{m},t}}=\frac{{\mathrm{P}_{\mathrm{mbs}}}}{10^{\xi_{{k,m}}(t)}/10}$. Similarly, the received power at user $k$ from  the UAV at time $t$, can be calculated as ${S_{{u,t}}}=\frac{{\mathrm{P}_{\mathrm{uav}}}}{10^{\xi_{{k,u}}(t)}/10}$. Here, $\xi_{{k,m}}(t)$ and $\xi_{{k,u}}(t)$ are path losses in dB which are calculated according to path loss models discussed in Section \ref{PL}. During each $t$, a UE connects to either its nearest MBS or the UAV, whichever provides the best signal-to-interference ratio (SIR). Assuming round-robin scheduling, we can express the spectral efficiency (bps/Hz) of user $k$ at time $t$ using Shannon's capacity  as: 
\begin{align}
{R_{k}(t)}=\frac{\log_{2}(1+{\gamma_{k}(t))}}{{N_{\mathrm{ue}}}},
\end{align}
where ${\gamma_{k}(t)}$  is the instantaneous SIR of $k$-th user at time $t$ and ${N_{\mathrm{ue}}}$ is the number of users in a cell. Let us assume the set of all transmitters of the network including the UAV as $\mathcal{X}$
. Then ${\gamma_{k}(t)}$ can expressed as:
\begin{align}
\gamma_{k}(t)=\frac{S_{{i},t}}{\sum_{j \neq i} S_{{j},t}},
\end{align}
where, $S_{{i},t}$ is the received power at user $k$ from  transmitter (MBS/ UAV) $i$,  with which the user $k$ is associated at time $t$. We normalize $R_{k}(t)$ by $T$ to get the time-averaged capacity of user $k$ and unless otherwise stated, the term `capacity' will refer to time-averaged capacity throughout this paper.
\begin{table}[t]
\label{tab}
\centering
\caption{\textbf{Simulation parameters}} 
\begin{tabular}{|c|c|} \hline
\textbf{Parameter} & \textbf{Value} \\ \hline
P\textsubscript{mbs} & 46 dBm  \\ \hline
P\textsubscript{uav} & 30 dBm \\ \hline
V\textsubscript{max} & 17.7 m/s\\ \hline

[$x(s),y(s)$] & [0, 0] km \\ \hline
[$x(f),y(f)$] & [1, 1] km\\ \hline
h\textsubscript{uav} & 120 meter\\ \hline
h\textsubscript{bs} & 30 meter\\ \hline
h\textsubscript{ue} & 2 meter\\ \hline
${f_c}$ & 1.5 GHz\\ \hline
$\mathrm{\alpha_L}, \mathrm{\alpha_N}$ & 2.09, 3.75\\ \hline
$\lambda$\textsubscript{mbs} & 2, 3,  4 per sq. km\\ \hline
$\lambda$\textsubscript{ue} & 100 per sq. km\\ \hline
\end{tabular}
\end{table}

\subsection{Simulation Assumptions}
In the rest of the paper, we numerically obtain the optimal trajectories of the UAV by applying DP and use simulations to analyze the effects of different network design aspects on the optimal trajectories of the UAV. The UEs and the MBSs are distributed in an area of 1 km $\times$ 1 km, whereas the UAV can fly on an area of 1.2 km $\times$ 1.2 km. This allows us to study the UAV behavior for different design parameters such as antenna radiation and backhaul constraint more explicitly. The results provide interesting insights on deploying mission time constrained drones as supportive network nodes for a cellular network. Unless otherwise specified, the simulation parameters and their default values are listed in Table I. They are chosen in a way to reflect a UAV traversing over an interference limited realistic downlink cellular network situated in a suburban area.

\begin{figure}
		\centering
		\subfloat[]{
			\includegraphics[width=0.92\linewidth]{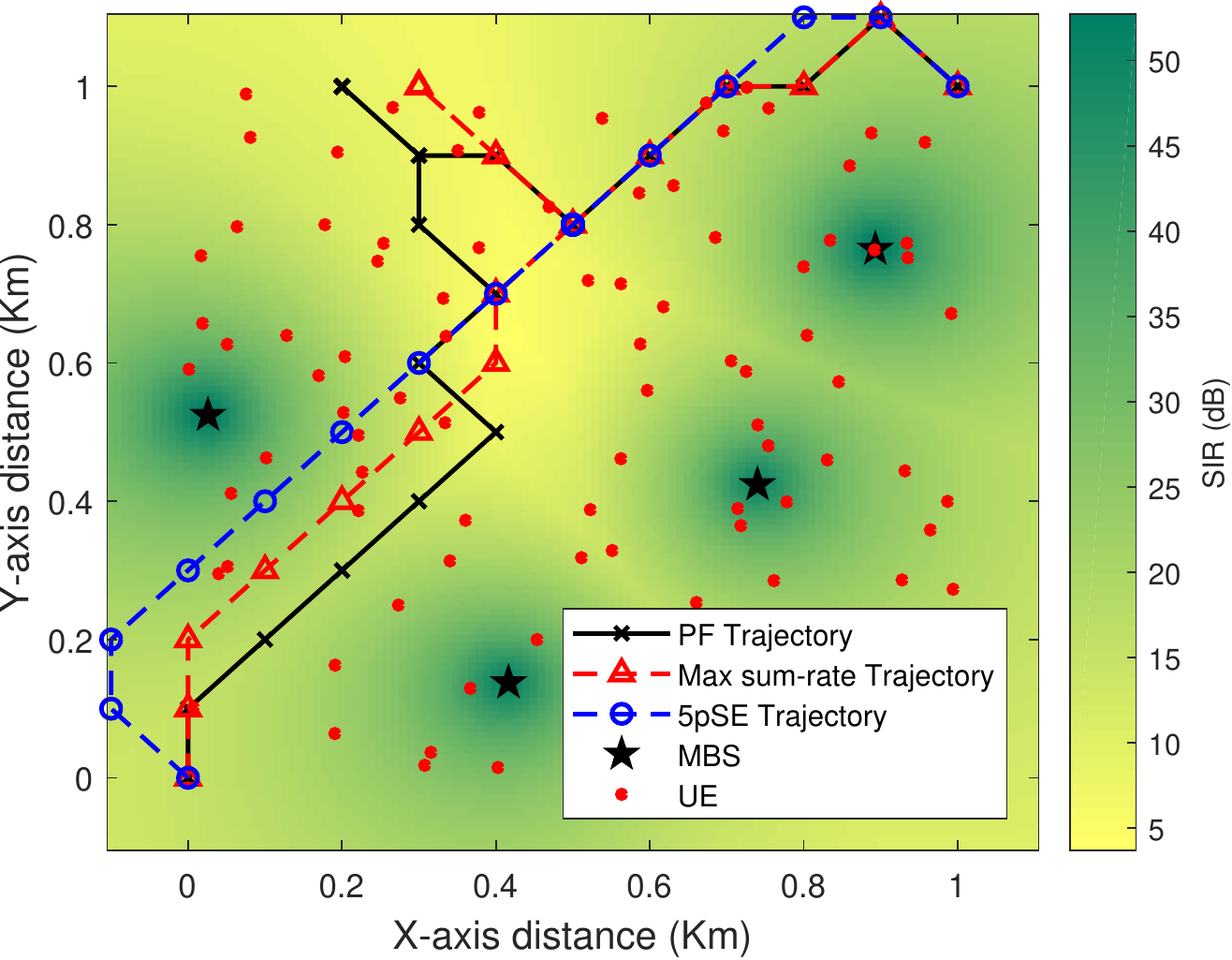}}
			
		\subfloat[]{
			\includegraphics[width=0.92\linewidth]{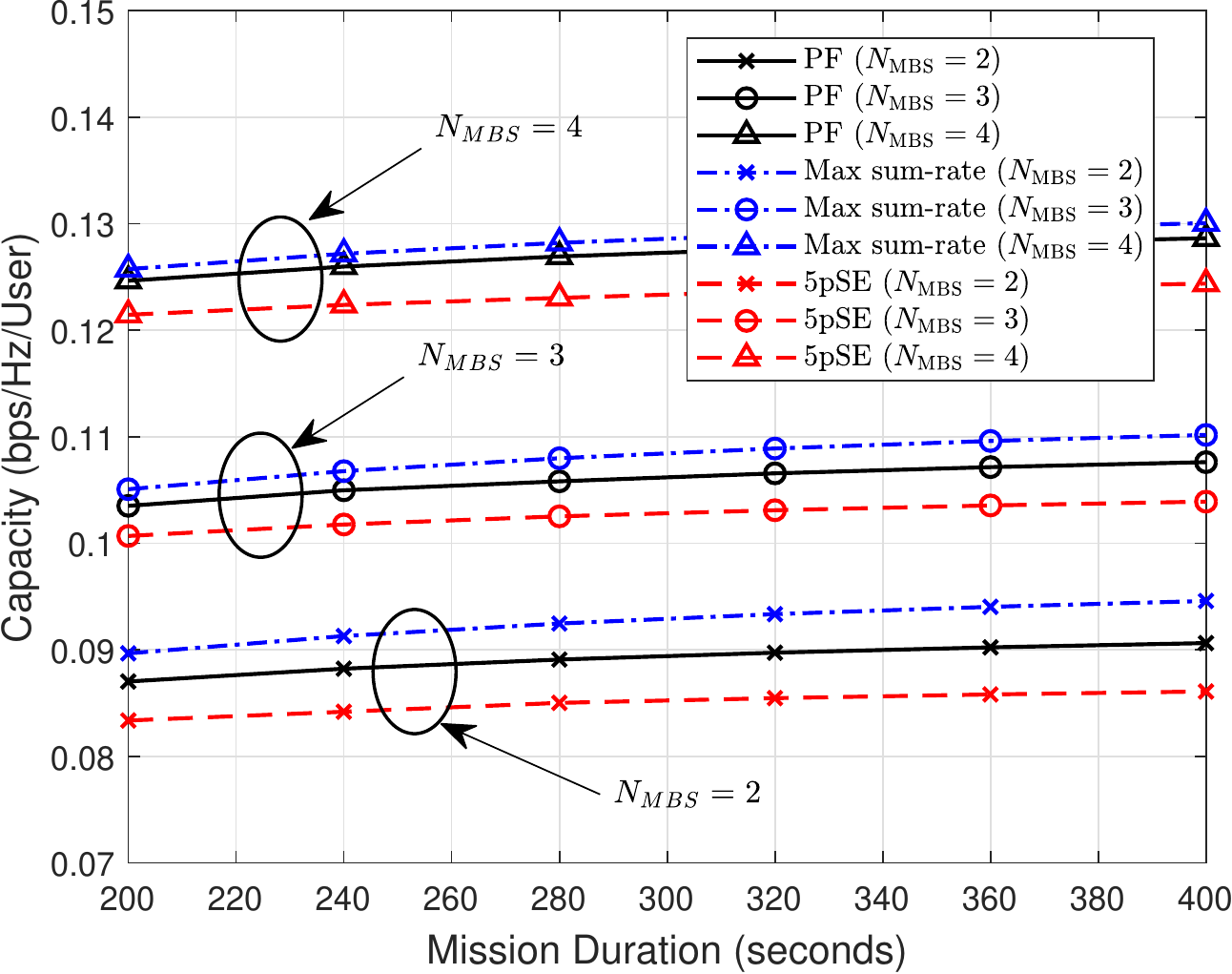}}
           		
		\subfloat[]{
			\includegraphics[width=0.92\linewidth]{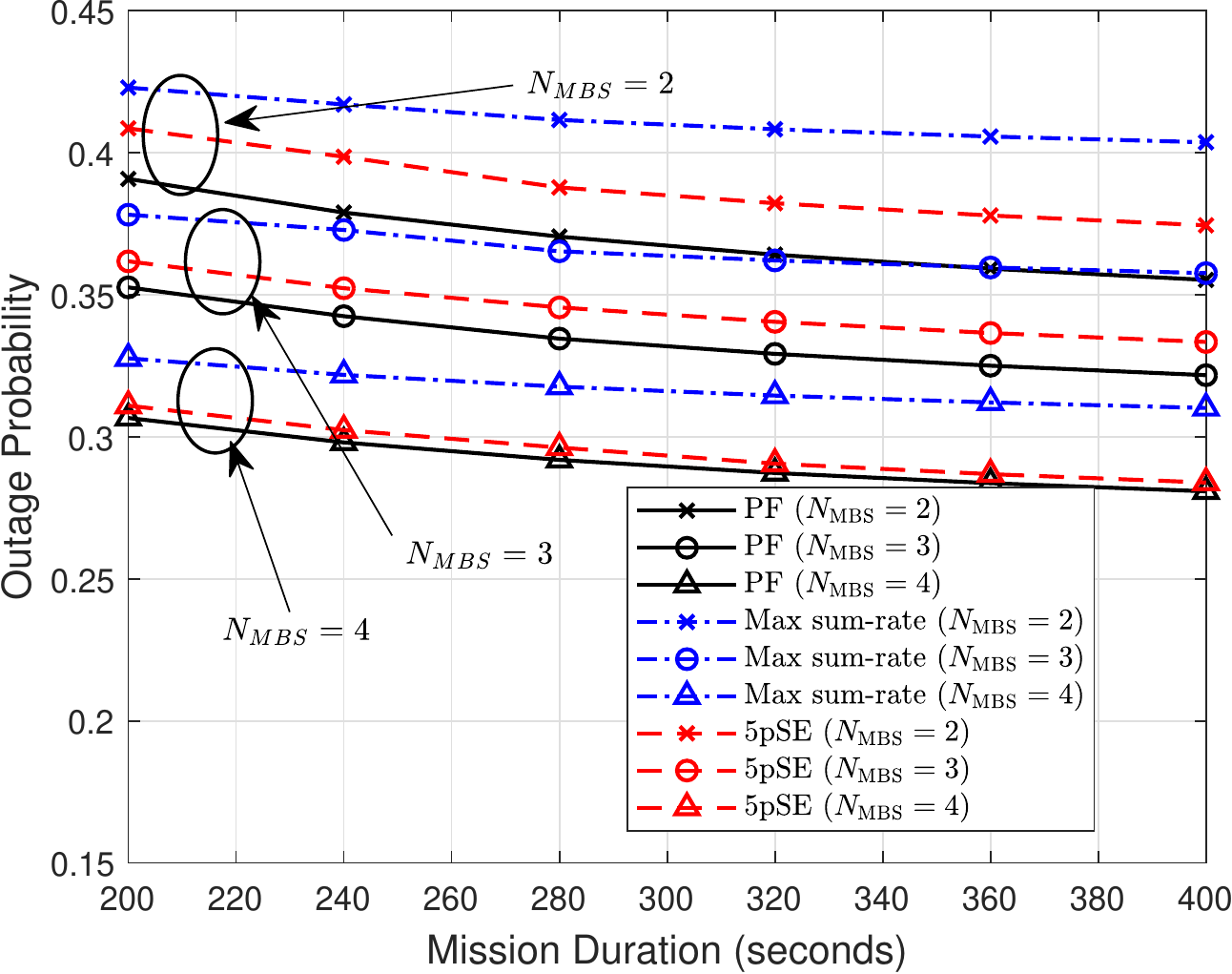}}
			
		\caption {\textbf{(a) Optimal trajectories of PF rate, Max sum-rate and 5pSE rate for \textbf{$T=240$}~s overlapped on SIR (dB) heat map at each discrete point.
		(b) Network per-user capacity comparison between PF rate, max sum-rate, and 5pSE trajectories. 
       (c) Outage Probability comparison between PF rate, max sum-rate, and 5pSE trajectories. 
        }}
		\label{pf_cap_sch}
	\end{figure}
To reduce the computational complexity, the possible UAV locations or states, the time, and the actions are divided into discrete segments. We discretize the square map $\mathcal{A}$, over which the UAV is allowed to move, into steps of 100 m, which results in 169 unique geometric positions. We also discretize the time into $\delta=$8 seconds. As we have segmented all possible states into finite discrete geometrical positions, we consider the following control actions on the map:
\begin{equation}
\label{act}
    \mathbf{u} \in \Bigg\{\begin{bmatrix} 0 \frac{m}{s}\\ \\0 \end{bmatrix},\begin{bmatrix} 12.5 \frac{m}{s}\\ \\\theta \end{bmatrix},\begin{bmatrix} 17.7 \frac{m}{s}\\ \\\theta+\frac{\pi}{4} \end{bmatrix} \Bigg\},
\end{equation}
where, $\theta \in \{0,\frac{\pi}{2},\pi,\frac{3\pi}{2} \}$. Actions in (\ref{act}) imply that the UAV either stays at it's current location or moves towards one of the eight directions separated by 45\degree increments.

\section{Impact of Scheduling Criteria}
In this section, we first consider the use of different scheduling criteria while assigning UEs to the MBSs or to the UAV. We consider proportional rate to be our baseline scheduler. We are also interested in studying how optimal trajectory behaves with other scheduling approaches such as, maximum sum-rate scheduler and fifth percentile rate scheduler (5pSE). The 5pSE scheduler takes the worst fifth percentile UE capacity of the network into account. In other words, 5pSE helps to maintain a minimum QoS level for all UEs of the network. For finding the optimal trajectory that corresponds to the maximum sum-rate, we can just exclude the $\log_{10}$() term in the optimization problem \eqref{eq}. In order to find the 5pSE optimal trajectory we can replace \eqref{Eq3} with:
\begin{maxi!}	
	  {x(t),y(t)}{\int_{t=0}^{T}C_{\textsubscript{5th}}(t)}{}{},\label{Eq4}
\end{maxi!}

where, $C_{\textsubscript{5th}}(t)$ represtens the 5th percentile user capacity across the whole network.

\begin{figure}\label{bez_example}
\centering
\includegraphics[width=0.92\linewidth]{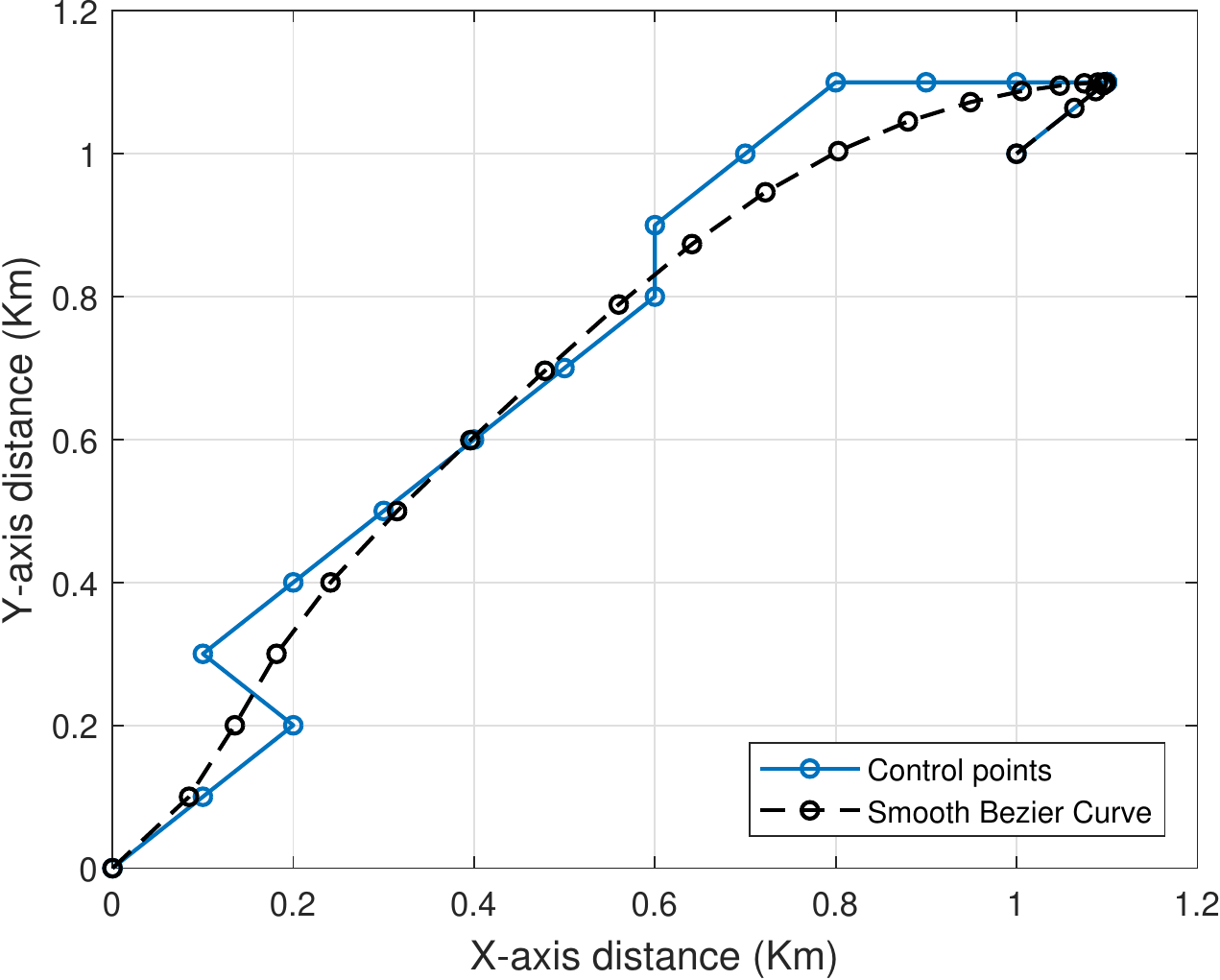}\\
\caption{\textbf{An example of path smoothing using Bezier curves with 25 control
points}}
\end{figure}

In the following, we investigate the trajectories associated with PF rate, maximum sum-rate, and 5pSE rate, in a network with 4 MBSs and 100 UEs for $T=240$~s, starting from source $(0,0)$~km to destination $(1,1)$~km. In Fig.~\ref{pf_cap_sch}(a), we plot optimal trajectories associated with three different schedulers, from which it is evident that the optimal paths are highly dependent on the signal-to-interference ratio (SIR) at each discrete point. The maximum sum-rate trajectory tends to move towards low SIR regions and tries to associate two or three UEs and provide downlink coverage to them. The PF rate trajectory tends to move in both high and low SIR regions to maintain a balance between rate and fairness. The 5pSE trajectory on the other hand, offloads some UEs with good throughput from the MBS to the UAV, so that, the total resource of the MBS can be distributed among fewer UEs which helps to improve the capacity of cell-edge UEs. Another interesting observation is that, while completing mission, the UAV tends to reach the optimal point (highest value among the 169 points) with V\textsubscript{max} and hover there for a while before it starts moving towards the final destination again with V\textsubscript{max} to meet the time constraint for all three approaches. These points are (0.2, 1)~km, (0.3, 1)~km and (0.5, 0.8)~km for PF, maximum sum-rate and 5pSE, respectively. 

\begin{figure}
		\centering
		\subfloat[]{
			\includegraphics[width=0.92\linewidth]{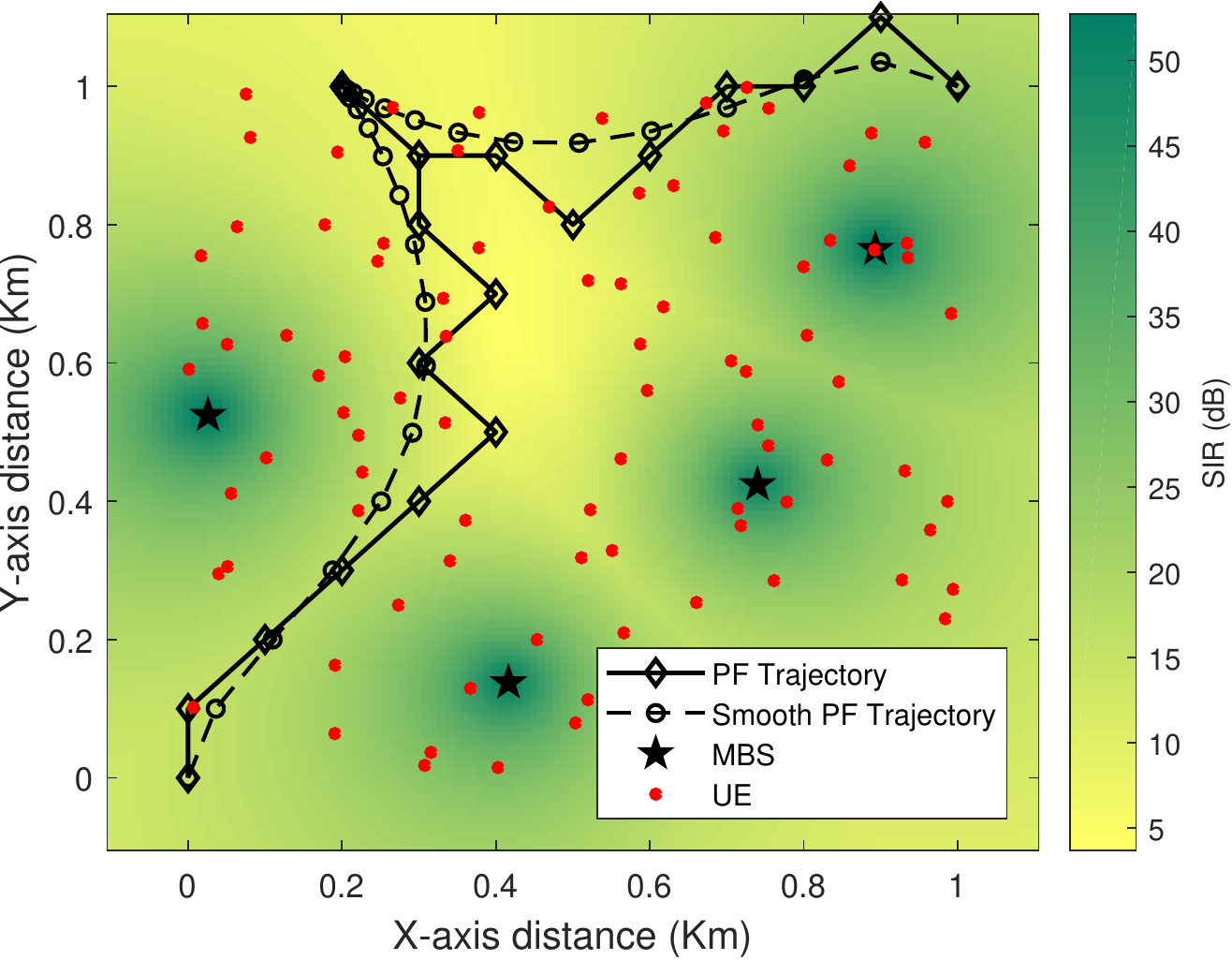}}
			
		\subfloat[]{
			\includegraphics[width=0.92\linewidth]{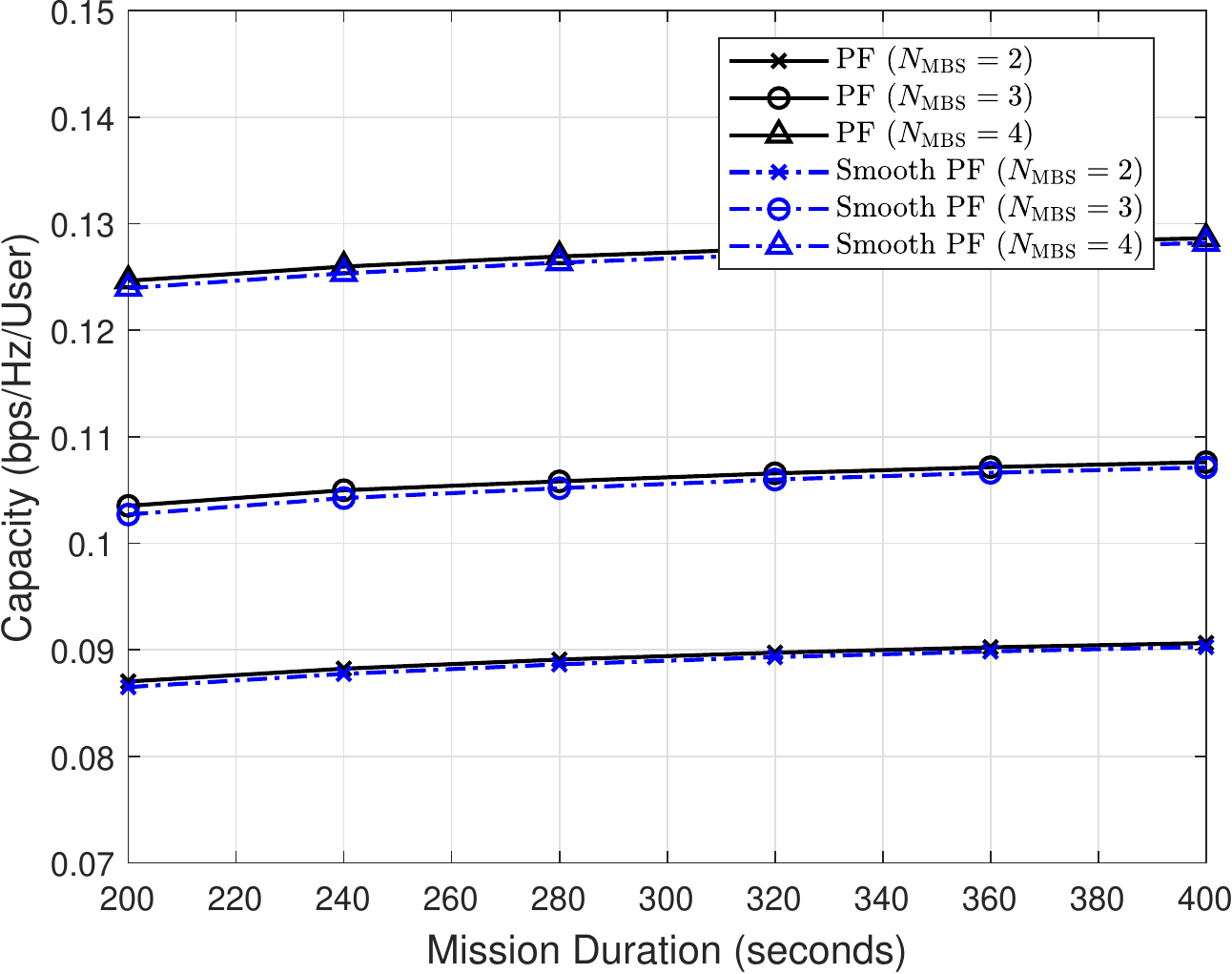}}
          		
		\subfloat[]{
			\includegraphics[width=0.92\linewidth]{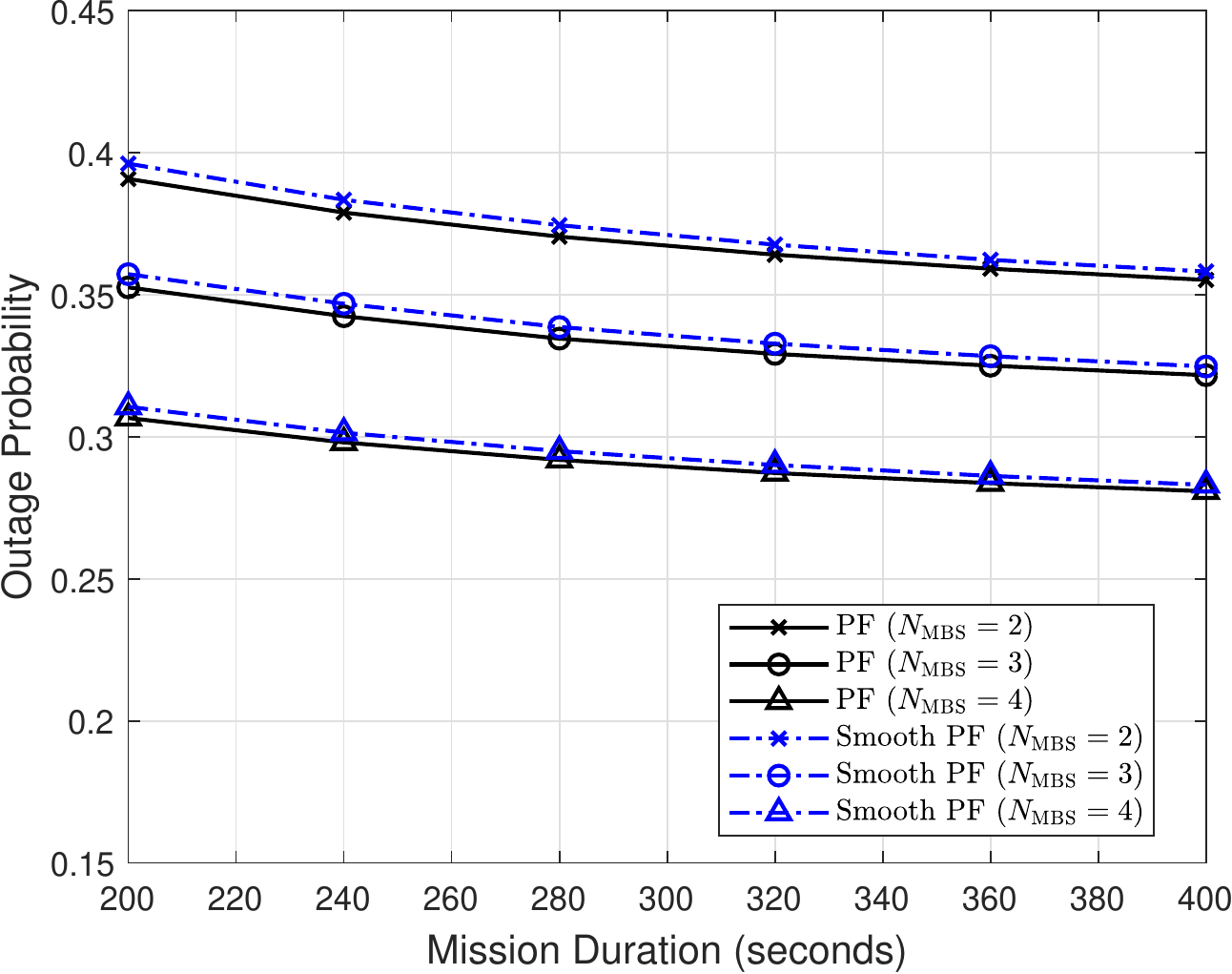}}
		\caption {\textbf{(a) Optimal trajectories of PF rate and smoothened PF-rate rate using bezier curve for \textbf{$T=240$}~s overlapped on SIR (dB) heat map at each discrete point. (b) Network per-user capacity comparison between PF rate, and smooth Pf trajectory generated by bezier curve. 
       (c) Outage Probability comparison between PF rate, and smooth PF trajectory generated by bezier curve.} 
        }
		\label{pf_cap_bez}
	\end{figure}
	
Next, we explore the capacity  performance of different trajectories by varying mission duration $T$ and number of MBSs ($N_{\mathrm{MBS}}$) in the network. We generated 30 different realizations of UE and MBS locations and calculated optimal trajectories for each $T$ and for various $N_{\mathrm{MBS}}$. Using the generated trajectories, we calculate the time averaged capacity per UE and average outage probability for all 30 networks. For the rest of the paper, we represent capacity and outage probability in the same way. Fig.~\ref{pf_cap_sch}(b) illustrates the time-averaged per-UE capacity for different criteria. With the increasing mission duration, normalized per-UE capacity increases due to the fact that UAV can reach out to further locations, and saturates for large $T$ as expected. The maximum sum-rate trajectory outweighs the other two approaches in terms of per-user capacity. Higher $N_{\mathrm{MBS}}$ results in better SIR and hence, provides better per UE capacity which is also evident from Fig.~\ref{pf_cap_sch}(a). Fig.~\ref{pf_cap_sch}(c) depicts the outage probabilities associated with the different trajectories, where we consider that a UE is in outage if its rate is lower than $0.05$~bps/Hz. For this scenario, the PF trajectory provides better coverage probability than max-sum rate and 5pSE rate. Outage probability decreases with the increasing number of MBSs due to better coverage and SIR. The maximum sum-rate trajectory does not take deprived UEs into account and hence, provides the worst performance.

\section{Impact of UAV Mobility Constraint}

In our simulations, paths are determined using dynamic programming algorithm by dividing time and the network area into discrete segments. However, these paths contain straight-line segments and sharp 90 degree turns. Generally they cannot be followed by UAVs due to their kinematic and dynamic constraints\cite{bezier1}. In fact, the UAV does not need to fly precisely over each points, rather it can move close to those points to avoid sharp turns. We therefore use Bezier curves to smooth the generated paths from dynamic programming technique. An illustration of the Bezier curve using 25 control points is presented at Fig.~\ref{bez_example}.
A Bezier curve is a smoothing curve in 2D space which uses Bernstein polynomials to generate the basis. A Bezier curve of degree $n$ or order $n+1$ can be written as\cite{bezier2}:

\begin{equation}
    b(t)=\sum_{i=0}^n \mathrm{\mathbf{P}_i}B_{i,n}(t),\quad 0\leq t \leq 1,
\end{equation}

where, $\textbf{P}$ is the vector of $n+1$ control points and $B_{i,n}(t)$ are the Bernstein polynomials of degree $n$ which can be explicitly expressed as:
\begin{equation}
    B_{i,n}(t)= {n \choose i} (1-t)^{n-i}t^i.
\end{equation}

The trajectories associated with PF rate and their corresponding smooth version generated from the Bezier curves are shown in Fig.~\ref{pf_cap_bez}(a). The smooth trajectory follows the original optimal path closely and generated the same number of control points as the original one. One interesting observation is, the UAV tends to move very slowly near the optimal point to maintain the performance.

			
          		
In Fig.~\ref{pf_cap_bez}(b) and Fig.~\ref{pf_cap_bez}(c), we plot the capacity and outage performance comparison of PF rate trajectory and its pertinent smooth trajectory. The analysis shows that the performance gap between these two trajectories is very small. This is due to the fact that, the Bezier curve allows the trajectory to remain almost stalled near the optimal point, which helps the UAV to maintain good capacity and outage performance. In essence, we can generate discrete trajectories using DP algorithm and subsequently obtain smooth trajectories using the Bezier curves without any significant degradation in the network performance. For the following sections, we will consider trajectories smoothed by Bezier curves.    

\section{Impact of Path Loss Model}
\label{PL}
One of our main goals of this research work is to study the effects of different path loss models on the optimal UAV trajectories. The three path loss models of our interest are discussed in the following subsections. Here, we consider sub-6 GHz band for downlink cellular network  which is interference limited. In other words, the presence of thermal noise power at a receiver is presumed to be negligible compared to the interference power. The Doppler spread stemming from the UAV's mobility is considered to be taken care of at the receivers. 
\subsection{Okumura-Hata Path Loss Model}
We consider Okumura-Hata path loss model (OHPLM) for MBS-to-UE links, as it is more relevant for a terrestrial environment where base-station height does not vary significantly~\cite{hata}. We also consider this model as our baseline path loss model for UAV-UE links.  
The path loss (in dB) observed at UE $k \in \mathcal{K}$ from MBS $m$ at time $t$ is given by:
\begin{equation}
 \xi_{{k,m}}(t)=A+B\log_{10}({d_{{k,m,t}}})+C,  
\end{equation}
 and path loss (in dB) observed at UE $k \in \mathcal{K}$ from the UAV at time $t$ is:
 \begin{equation}
    \xi_{{k,u}}(t)=A+B\log_{10}({d_{{k,u,t}}})+C. 
 \end{equation}
 Here, ${d_{{k,m,t}}}$  and ${d_{{k,u,t}}}$ are the Euclidean distances from MBS $m$ to user $k$ and from UAV to user $k$ at time $t$. $A$, $B$, and $C$ are the factors dependent of the carrier frequency $f_c$ and antenna heights \cite{hata}. The expressions of the factors $A$, $B$, and $C$, in a suburban environment are given by ~\cite{hata}:
 \begin{equation}
A=69.55+26.16\log_{10}(f_c)-13.82\log_{10}(\mathrm{h}_{\mathrm{bs}})-a(\mathrm{h}_{\mathrm{ue}}),\\
 \end{equation}
\begin{equation}
    B=44.9-6.55\log_{10}(\mathrm{h}_{\mathrm{bs}}),\\
\end{equation}
\begin{equation}
     C=-2\log_{10}(f_c/28)^2-5.4,
 \end{equation}
 where, $f_c$ is carrier frequency in MHz, $\mathrm{h}_{\mathrm{bs}}$ and $\mathrm{h}_{\mathrm{ue}}$ are the height of the base station (MBS/UAV) and the height of UE antenna in meter unit, respectively. The correction factor $a(\mathrm{h}_{\mathrm{ue}})$ due to UE antenna height can be defined as,
 \begin{equation}
     a(\mathrm{h}_{\mathrm{ue}})=[1.1\log_{10}(f_c)-0.7]\mathrm{h}_{\mathrm{ue}}-1.56\log_{10}(f_c)-0.8.
 \end{equation}

 The OHPLM assumes the carrier frequency ($f_c$) to be in the range between 150 MHz and 1500 MHz, $\mathrm{h}_{\mathrm{bs}}$  between 30 m to 200 m, and $\mathrm{h}_{\mathrm{ue}}$ between 1 m to 10 m. Furthermore, this model also assumes the distances $d_{\mathrm{k,m,t}}$ and $d_{\mathrm{k,u,t}}$ between 1 km to 10 km. Since, the highest possible distance in our simulations can be 1.414 km, this model is a good candidate for our study. 
 \subsection{Mixture of LOS/NLOS Path Loss Model}
 To study the effects of various path loss models between UAV-to-UE links on the optimal trajectories, here we deploy a mixture of LoS/NLoS propagation model (MPLM) for modeling the path loss between UAV and the UEs\cite{mixturepl}. As the UAV flies above the ground, it has a higher probability of LOS. On the other hand, due to the presence of man-made structures, the link between UAV and UE can be in NLOS.  This motivated us to study how this mixture model shapes the optimal UAV trajectories. Given a horizontal distance, $z_{{k,u,t}}$,  between a UE $k$ and the UAV at time $t$, the LOS probability $\tau_L(z_{{k,u,t}})$ can be defined as\cite{mixturepl}:
\begin{equation}
 \tau_L(z_{{k,u,t}})=\prod_{n=0}^{m} \Bigg( 1-\exp { \Bigg\{- \frac {\mathrm{h}_{\mathrm{uav}}-(n+0.5)(\mathrm{h}_{\mathrm{uav}}-\mathrm{h}_{\mathrm{ue}})}{2\hat{c}^2} \Bigg\}} \Bigg),
\end{equation}
where, $ m=\floor*{ \frac {z_{{k,u,t}}\sqrt{\hat{a}\hat{b}}}{1000}-1}$. Here, a suburban area is defined as a set of buildings placed in a square grid in which $\hat{a}$ stands for fraction of the total land area occupied by the buildings, $\hat{b}$ is the mean number of buildings per sq. km, and the buildings height is defined by a Rayleigh PDF with parameter $\hat{c}$\cite{mixturepl}. Consequently, the NLOS probability can be expressed as: 
\begin{equation}
    \tau_N(z_{{k,u,t}})=1-\tau_L(z_{{k,u,t}}).
\end{equation}
After calculating the LOS and NLOS probabilities, we can get the path loss (in dB) at UE $k \in \mathcal{K}$ from the UAV at time $t$ as:
\begin{equation}
\begin{split}
 \xi_{{k,u}}(t)= 10\log_{10}(\mathrm{P}_\textsubscript{uav} [(d_{k,u,t})^{-\alpha_{L}}\tau_L(z_{k,u,t})\\+(d_{{k,u,t}})^{-\alpha_{N}}\tau_N(z_{{k,u,t}})]),   
\end{split}
\end{equation}
where, $\mathrm{\alpha_{L}}$ and $\mathrm{\alpha_{N}}$ are the path loss exponents associated with LOS path and NLOS path, respectively. For our simulations, we considered values of $\hat{a}$,$\hat{b}$, and $\hat{c}$ to be 0.1, 100, and 10, respectively. Values of $\mathrm{\alpha_{L}}$ and $\mathrm{\alpha_{N}}$ are provided in Table 1.
\begin{figure}
		\centering
		\subfloat[]{
			\includegraphics[width=0.92\linewidth]{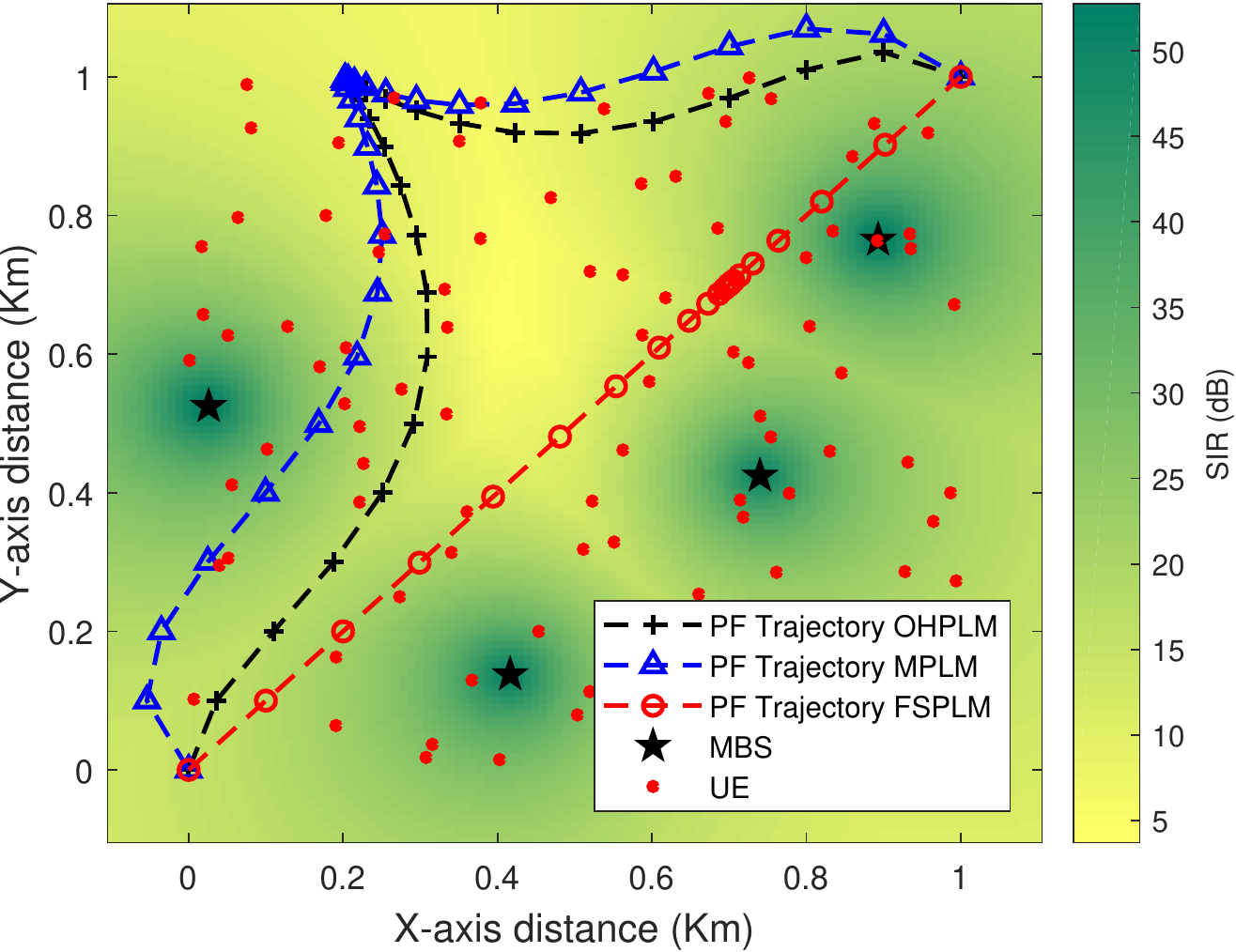}}
			
		\subfloat[]{
			\includegraphics[width=0.92\linewidth]{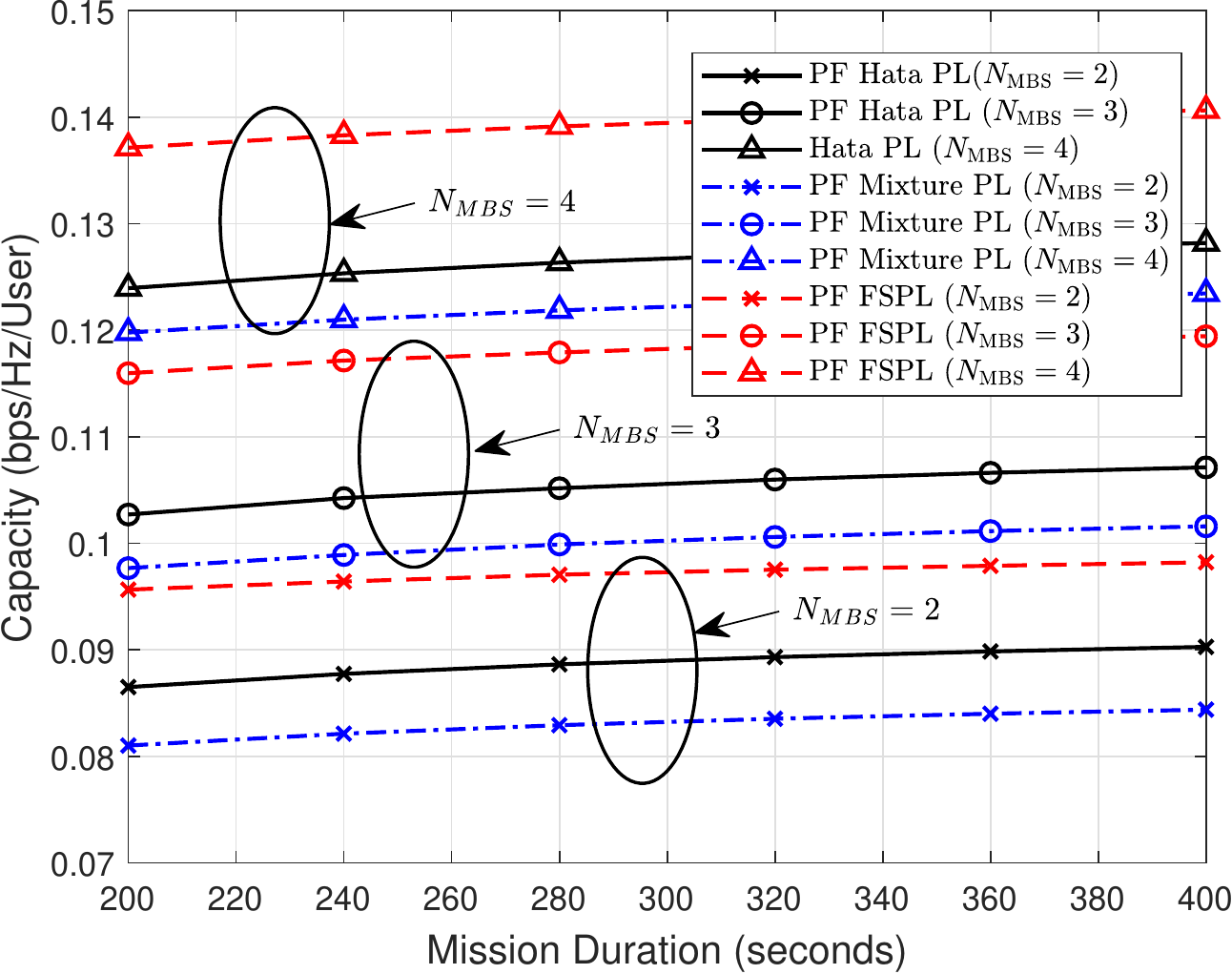}}
			
		\subfloat[]{
			\includegraphics[width=0.92\linewidth]{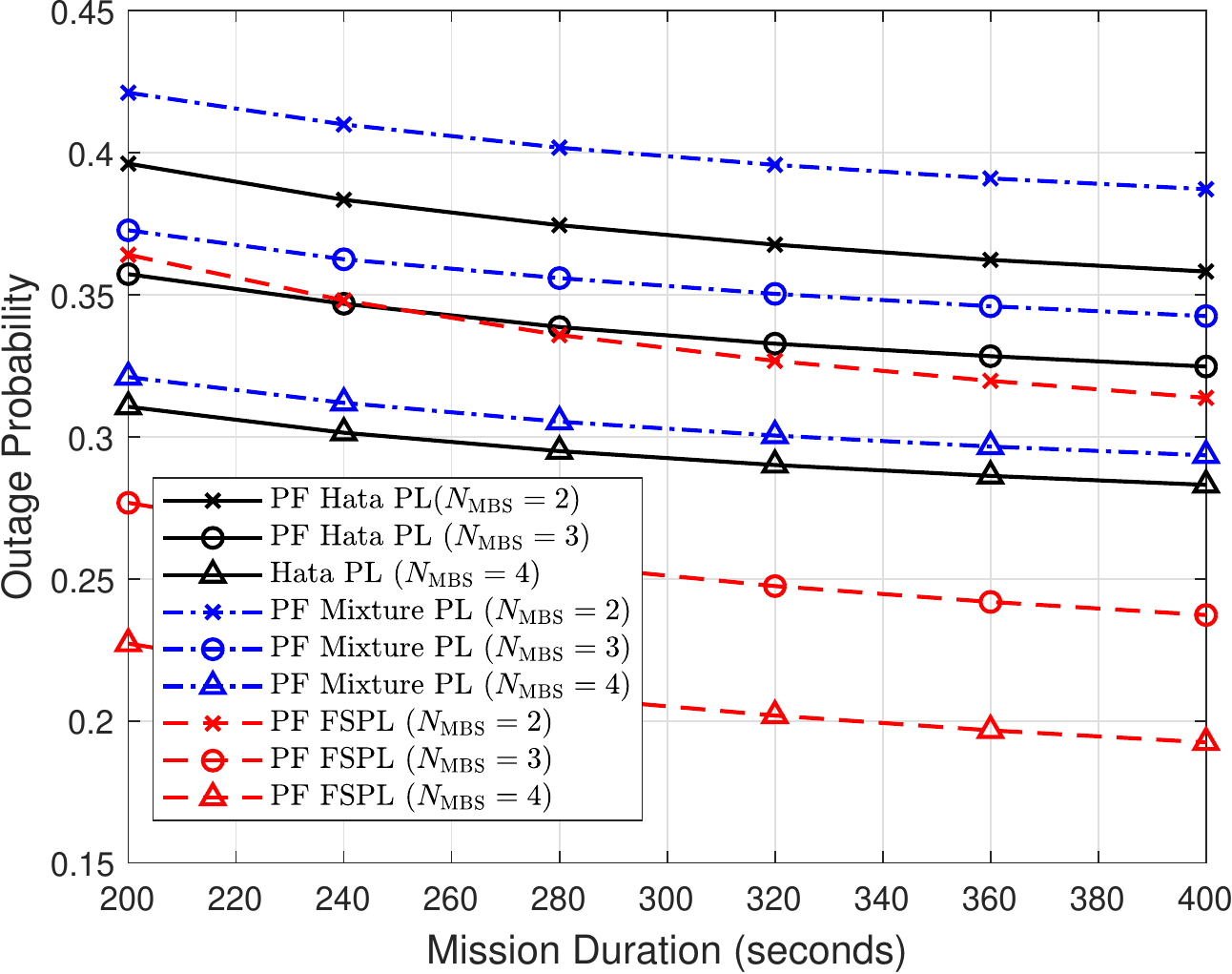}}
		\caption {\textbf{(a)  Optimal trajectories of PF rate associated with OHPLM, MPLM and FSPL for \textbf{$T=240$}~s overlapped on SIR (dB) heat map at each discrete point. (b) Network per-user capacity comparison in 3D antenna radiation effect between PF rate, max sum-rate, and 5pSE trajectories. 
       (c) Outage Probability comparison in 3D antenna radiation effect between PF rate, max sum-rate, and 5pSE trajectories. 
        }}
		\label{pf_cap_pl}
	\end{figure}
\subsection{Free Space Path Loss Model}
We also consider free space path loss model (FSPL), which is obtained from the Friis transmission equation\cite{fspl}. As UAVs are blessed with the advantage of getting LOS links with higher probabilities, we think FSPL model is a good candidate to explore. According to this model, the instantaneous path loss (in dB) observed at UE $k \in \mathcal{K}$ from the UAV at time $t$ can be defined as,
\begin{equation}
 \xi_{{k,u}}(t)= 20\log_{10}(d_{{k,u,t}}) + 20\log(f_c)- 27.55,  
\end{equation}
 where, $f_c$ is carrier frequency in MHz.
 
In the following, we focus our study on the effects of three different path loss models discussed in Section \ref{Sec2}. At first, in Fig. \ref{pf_cap_pl}(a), we plot optimal trajectories associated with three different path loss models, which also depicts the dependence of the optimal trajectories on the SIRs of the  discrete geometrical positions. The FSPL model provides around 8 fold more received power than its other two counterparts and hence, the UAV shows less tendency to hover over low SIR regions. The corresponding received power observed at UEs due to OHPLM and mixture model are close to each other which is also reflected by the similar trend of their relevant trajectories. 

Next, we explore the capacity and outage performance comparison between the trajectories associated with the three path loss models of our interest, which is illustrated in Fig.~\ref{pf_cap_pl}(b) and Fig.~\ref{pf_cap_pl}(c). As we mentioned above, the higher received power caused by FSPL model translates into higher SIR and subsequently into higher capacity and outage performances. The received power due to OHPLM is higher than that of MPLM and thus OHPLM outperforms Mixture path loss model in both capacity and outage performances. In the following sections, we will consider OHPLM for MBS-to-UE link and MPLM for UAV-to-UE link to calculate and compare the network performances.       
\begin{figure}\label{bachhaul}
\includegraphics[width=1.1\linewidth]{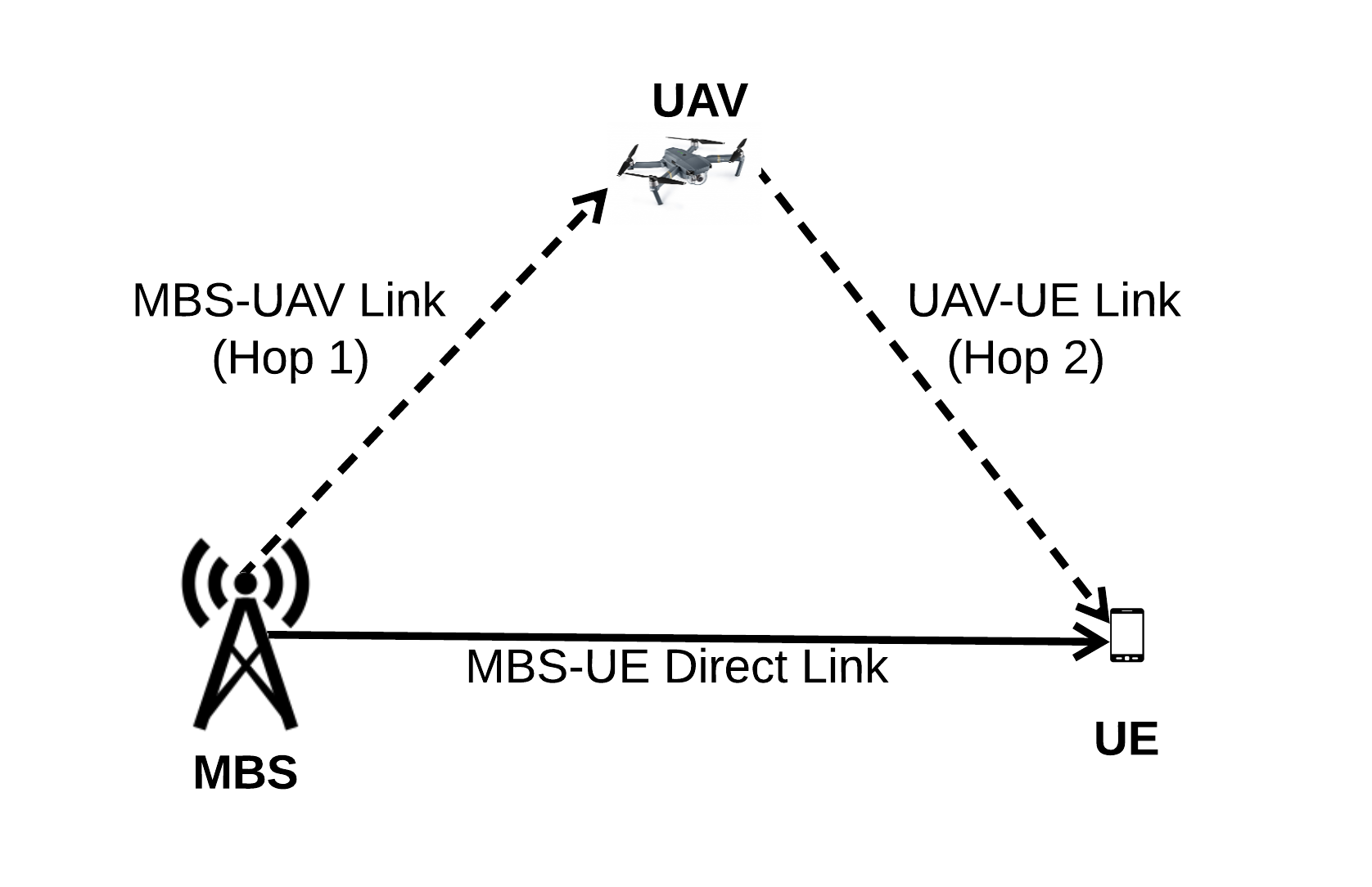}\\
\caption{\textbf{An illustration of using UAV as relay }}
\end{figure}

\begin{figure}
		\centering
		\subfloat[]{
			\includegraphics[width=0.92\linewidth]{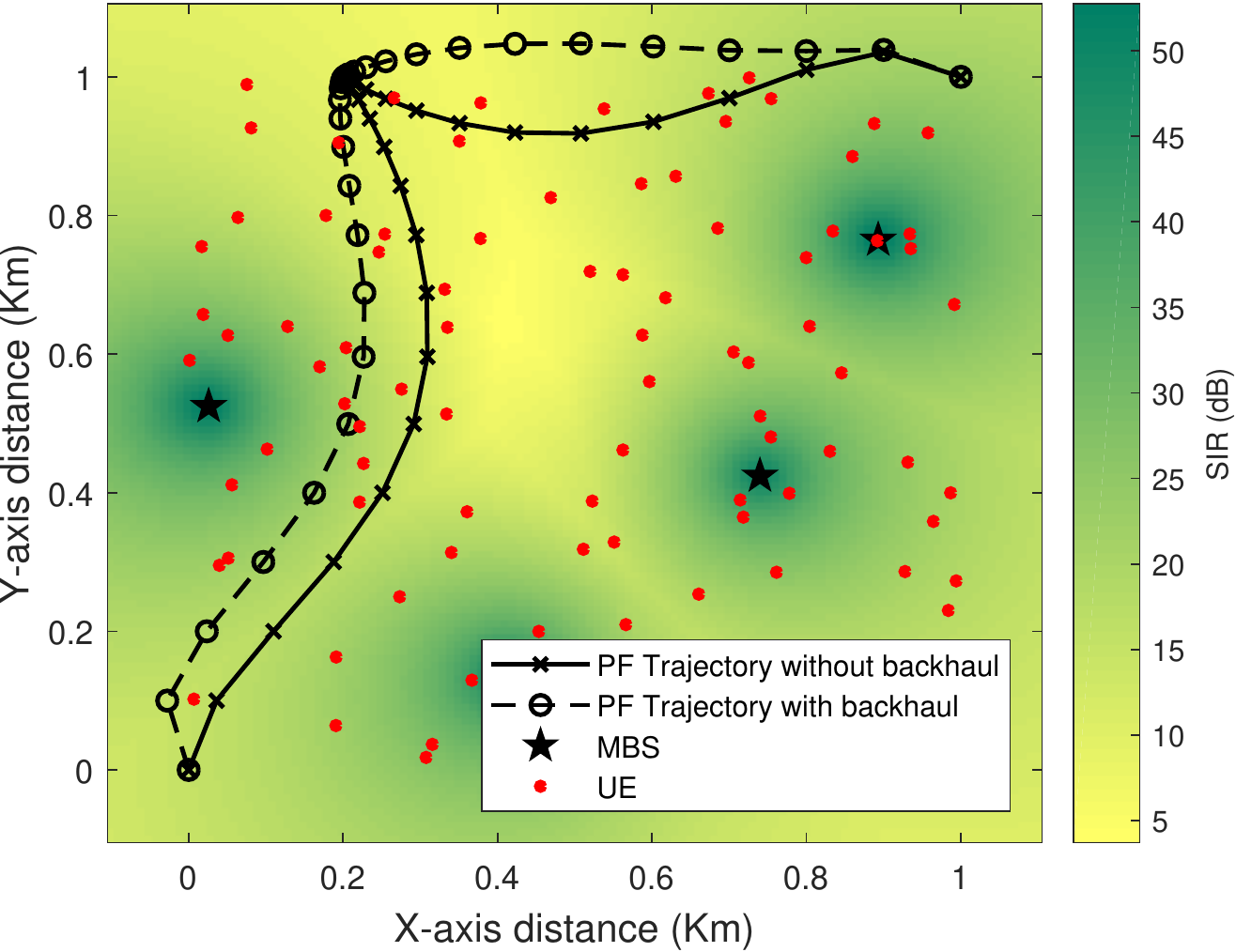}}
			
		\subfloat[]{
			\includegraphics[width=0.92\linewidth]{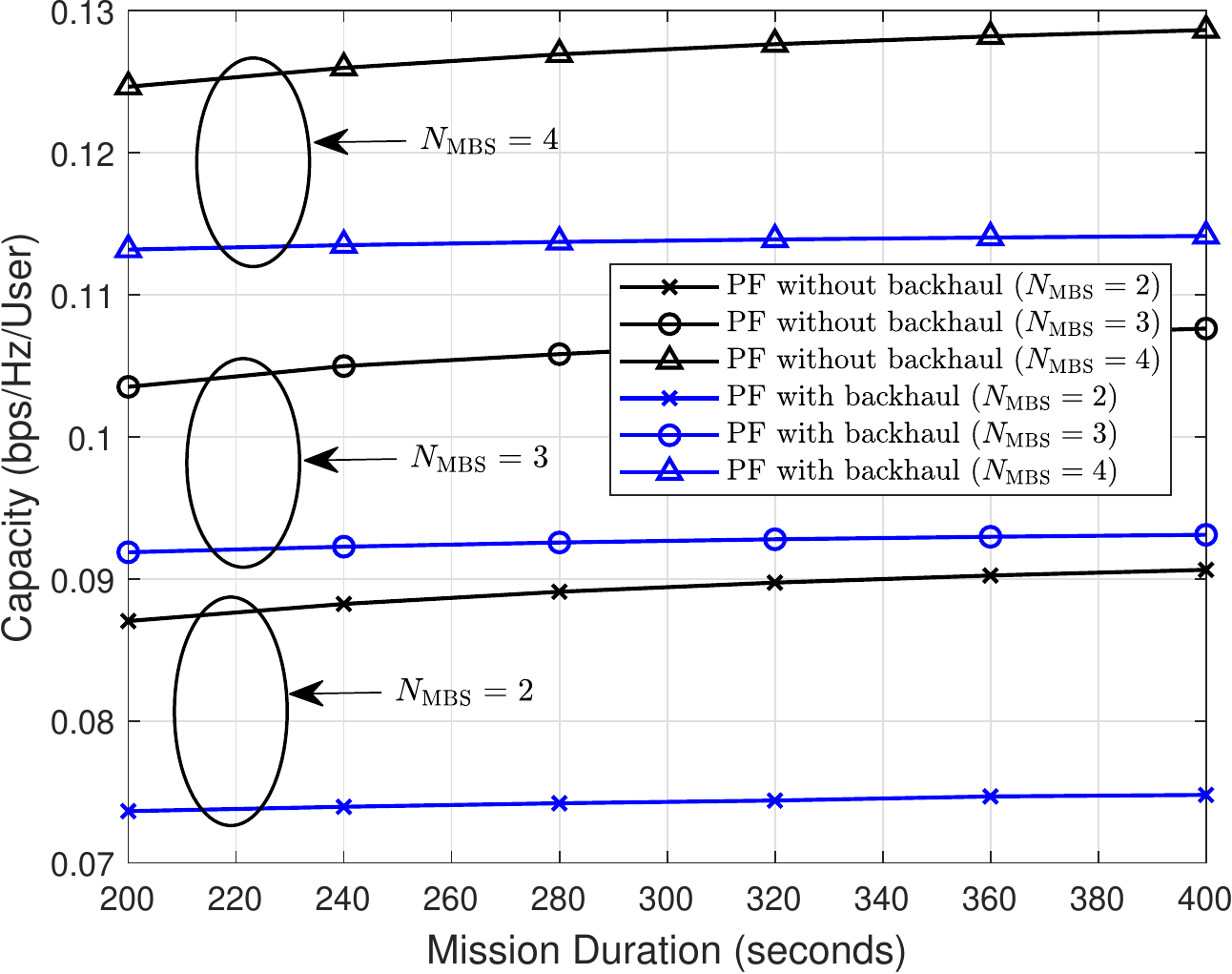}}
          		
		\subfloat[]{
			\includegraphics[width=0.92\linewidth]{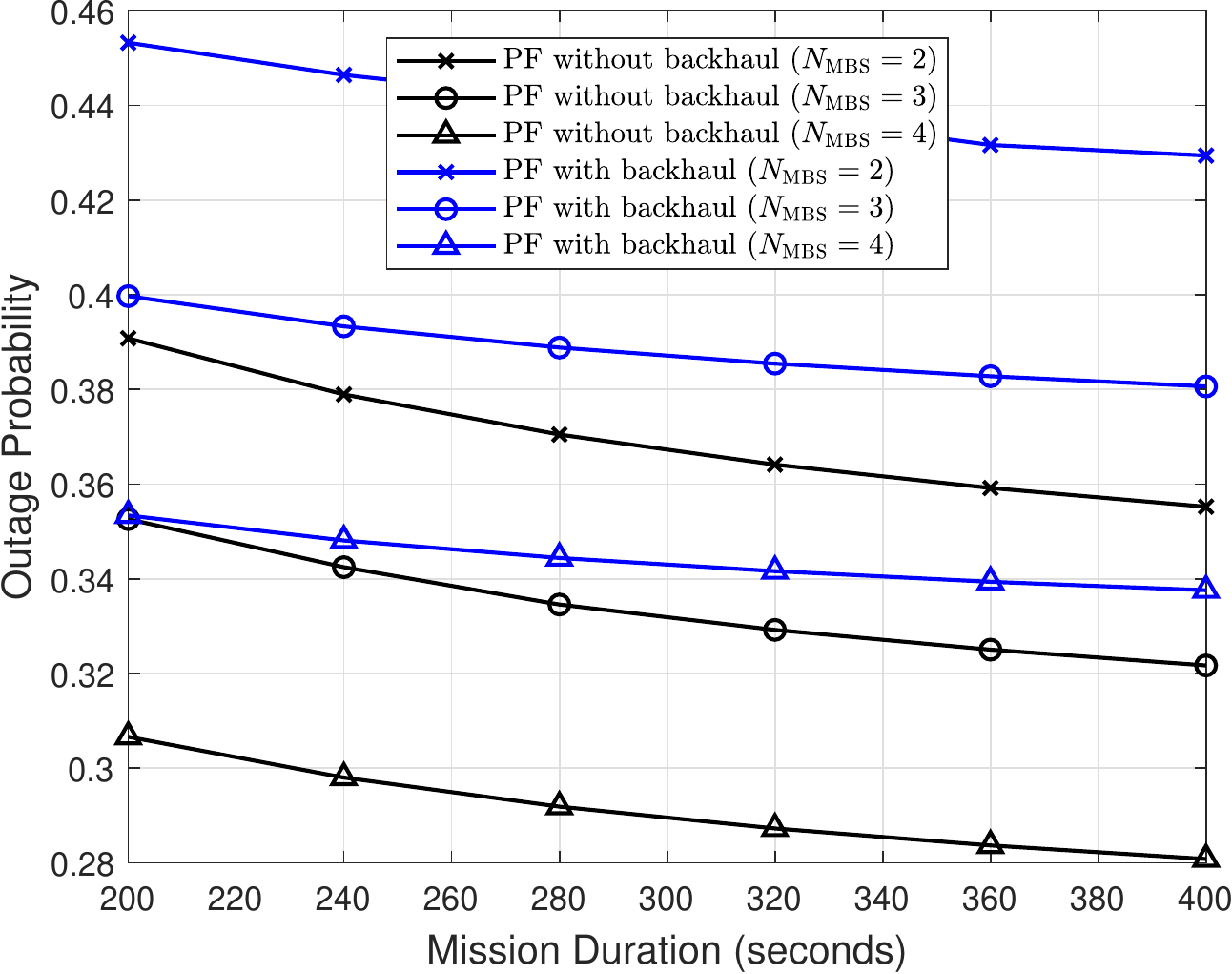}}
		\caption {\textbf{(a) Optimal trajectories of PF rate trajectories associated with and without backhaul constraint for \textbf{$T=240$}~s overlapped on SIR (dB) heat map at each discrete point. (b) Network per-user capacity comparison between PF trajectories associated with and without backhaul constraint. 
       (c) Outage Probability comparison between PF trajectories associated with and without backhaul constraint}. 
        }
		\label{pf_cap_back}
	\end{figure}	
	
\section{Impact of Backhaul Constraint}

Until this point, we have considered that, the UAV has somehow maintained constant connectivity with the backhaul. In fact, for reliable and safe mission critical UAV operation, the backhaul constraint via Command and Control (C2) link is a prerequisite\cite{backhaul}. This motivated us to consider the UAV as relay between the MBSs and the UEs in downlink and study the network performances. For the relay operation at the UAV, we consider amply and forward (AF) technique, where the end-to-end SIR is calculated as the harmonic mean of the SIRs related to UAV-UE link and MBS-UAV link\cite{chetan},\cite{sir}. An example of using UAV as a relay in downlink scenario is depicted in Fig.~\ref{bachhaul}. We consider 3GPP path loss model\cite{3gpp} for calculating the SIR for  MBS-UAV link. Let us assume, the SIRs associated with MBS-UAV link as $\gamma_\textsuperscript{mbs-uav}$ and $\gamma_\textsuperscript{uav-ue}$ for a UE $k$. According to \cite{sir}, the end-to-end SIR of UE $k$ can be calculated as:
\begin{equation}
    \gamma_k=\frac{2\gamma_\textsuperscript{mbs-uav} \times \gamma_\textsuperscript{uav-ue}}{\gamma_\textsuperscript{mbs-uav}+\gamma_\textsuperscript{uav-ue}}.
\end{equation}

If $\gamma_k > \gamma_\textsuperscript{mbs-uav} $, UE $k$ will be associated with the UAV and with the nearest MBS otherwise. In Fig.~\ref{pf_cap_back}(a), we plot the PF rate trajectories for both with and without backhaul constraint scenarios. As expected, to maintain the backhaul connectivity, the UAV tends to move closer to the MBSs. However, there is a catch since harmonic means are heavily influenced by smaller values. Therefore, in order to obtain a reasonable end-to-end SIR for overcoming the SIR stemmed from an MBS, the UAV will avoid going very close to the MBSs. Even though the MBS-to-UAV link SIR will be large if UAV gets closer to MBSs, the SIR pertaining to the UAV-to-UE link will be degraded severely due to strong interference from MBS. On the other hand, if the UAV moves too far away from the MBS to get closer to UEs, the opposite scenario will take place, i.e., the UAV-to-UE link SIR may be good but MBS-to-UAV link SIR  will degraded severely. Hence, the UAV will try to find UEs who are situated at low SIR region, while staying at regions where the SIR is not too high and not too low.

In Fig.~\ref{pf_cap_back}(b) and Fig.~\ref{pf_cap_back}(c), we plot the capacity and outage performance comparison of PF rate trajectory with and without backhaul connectivity constraint by using the path loss models mentioned above. It can be concluded that taking the backhaul constraint of the UAV into account will result in lower capacity as it can not associate users lying at low SIR regions and has less freedom to associate any UE lying on its trajectory. Apart from this, the UAV is actually acting as a UE from MBS point of view, i.e., the number of UEs associated to MBSs increases when the UAV acts as a relay. As a result, the UAV can only increase the network capacity to a small extent and hence, the time-averaged capacity performance does not change with mission duration as opposed to the previous results. The outage performance also suffers noticeable degradation when considering the backhaul constraint. \\

\begin{figure}
		\centering
		\subfloat[]{
			\includegraphics[width=0.92\linewidth]{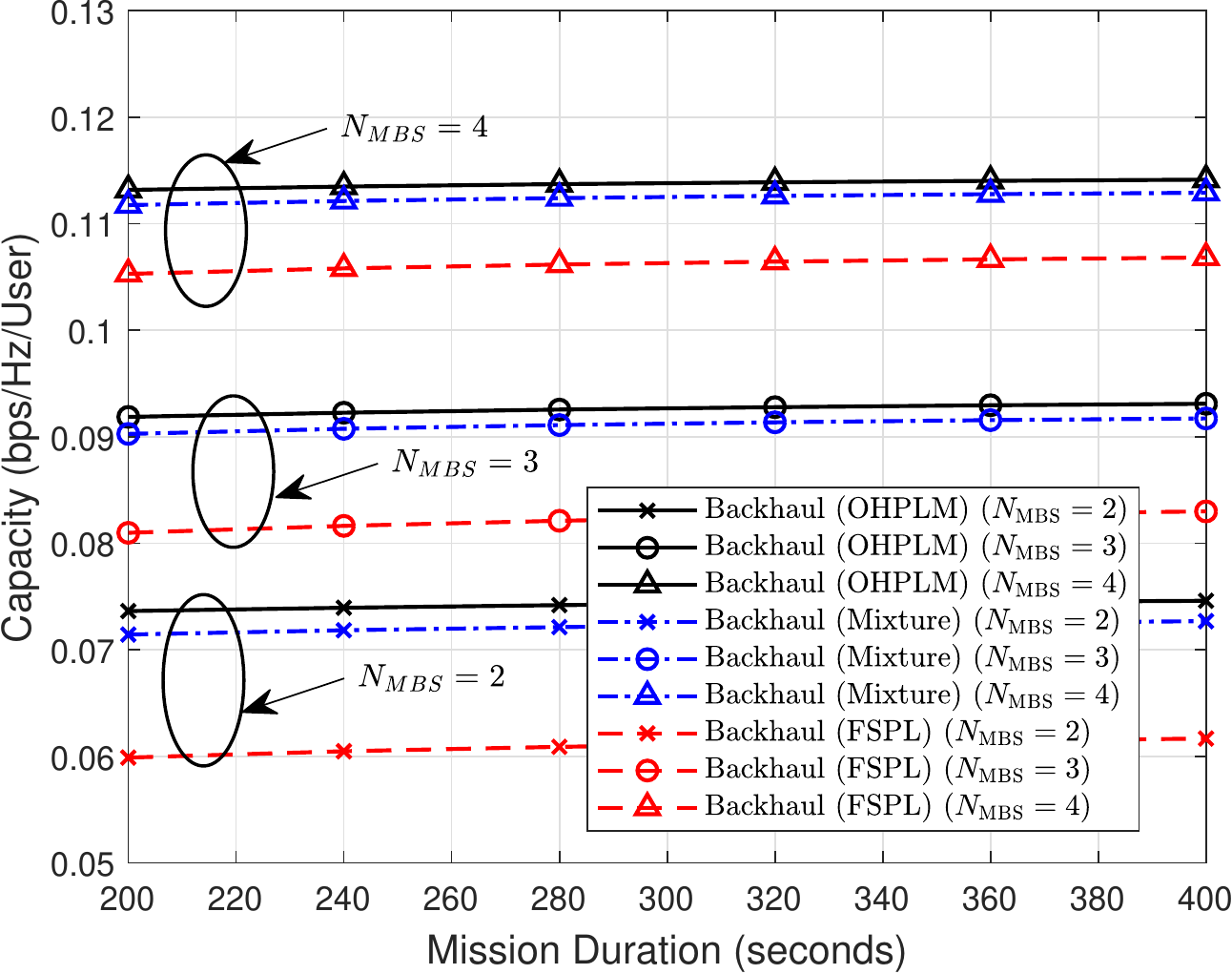}}
          		
		\subfloat[]{
			\includegraphics[width=0.92\linewidth]{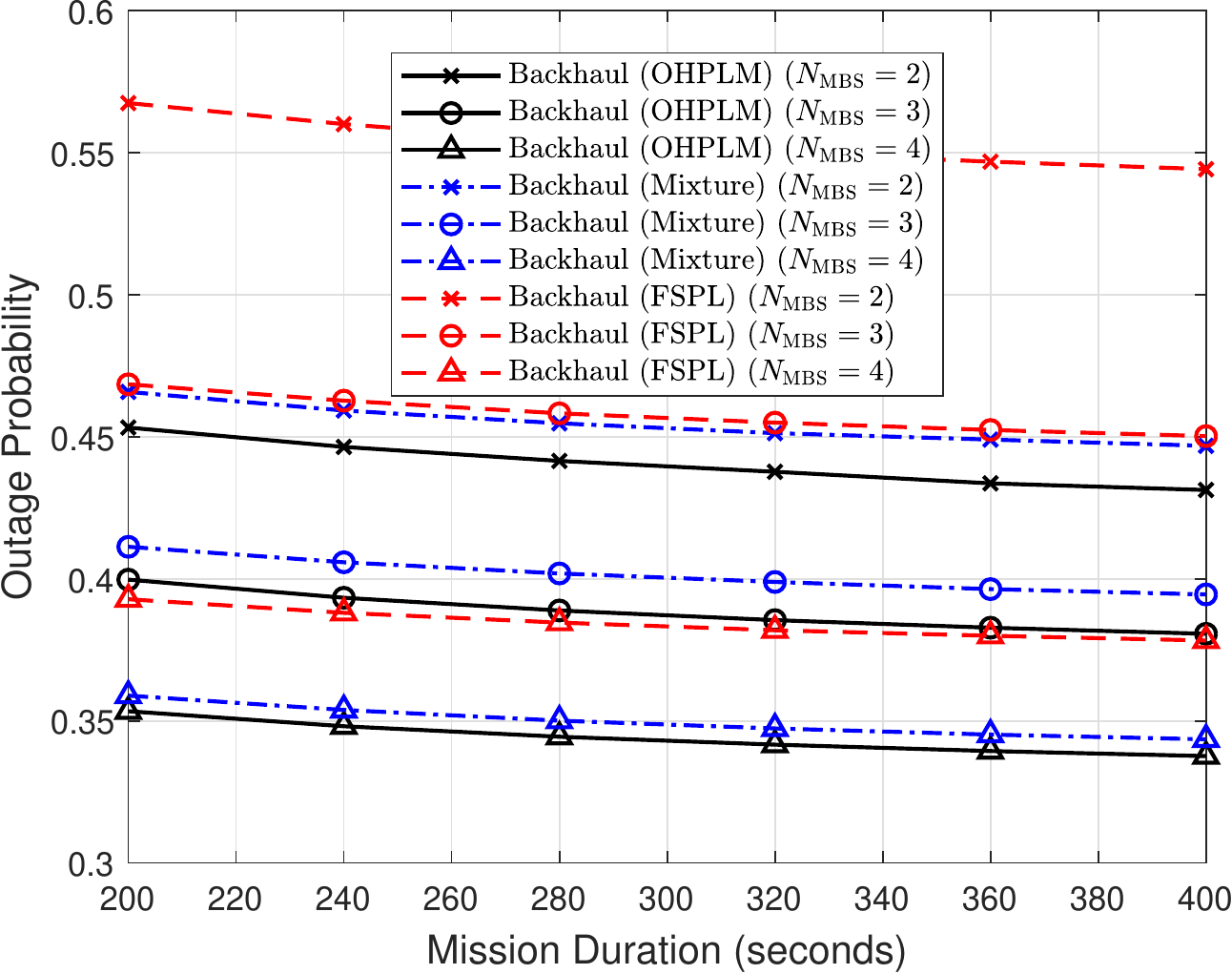}}
		\caption {\textbf{(a) Network per-user capacity comparison between trajectories with backhaul constraints for different path loss models. 
       (b) Outage Probability comparison between trajectories with backhaul constraints for different path loss models. }. 
        }
		\label{pf_cap_back2}
	\end{figure}

We also explore the effects of different path loss models on the optimal trajectories associated with backhaul constraints. In Fig.~\ref{pf_cap_back2}(a) and Fig.~\ref{pf_cap_back2}(b), we plot the capacity and outage performances for different path loss models. Here, we can see that the FSPL model provides the worst capacity and outage performances. This is due to the fact that the FSPLM provides higher received power for UAV-UE link. Hence, to maintain a balance between SIRs associated with MBS-UAV link and UAV-UE link, the UAV will move far from MBS which in turns degrades the end-to-end SIR. On the other hand, OHPLM and MPLM provides very close path losses for UAV-UE link, and hence, their outage and time average capacity performance do not have significant difference. The OHPLM provides slightly better performance due to its lower path loss than that of the MPLM.

\section{Impact of 3D Antenna Radiation}
Three dimensional (3D) radiation pattern of an antenna mounted at a UAV can significantly influence the air-to-ground (A2G) link quality\cite{priyanka_3d}. Even when a UAV hovers over a UE, if the antenna orientations and polarization are not aligned properly, an unexpected signal quality degradation can be observed at the receiver\cite{mmimo}. 

So far, we have considered omnidirectional antennas mounted on the UAV, UEs, and the MBSs. To study how antenna radiation pattern and polarization affect optimal trajectory path planning, we assume that both the UAV and the MBSs are equipped with two orthogonally crossed dipole antennas (one dipole antenna is oriented to z-axis and the other to the y-axis). For simplicity, we assume UEs are equipped with omnidirectional antennas. The polarization of an antenna can be defined as the direction of the transmitted waves radiated fields, evaluated at a given point in the far field\cite{Balanis}. Since, circularly polarized antenna suffers from lower polarization loss\cite{mmimo}, we consider the polarization of all antennas at UAV and the MBSs to be circular.  

In case of backhaul constraint, we model the combined gain of polarization loss factor (PLF) and antenna gain at the UAV for the MBS-UAV link using \cite[eq. (16)]{mmimo}. For the UAV-UE link and the MBS-UE link, we consider polarization factor stemming only from the transmitter side. After multiplying these gains with the corresponding received powers, we can envisage the effects of antenna radiation pattern on optimal UAV paths.

\begin{figure}
		\centering
			\subfloat[]{
			\includegraphics[width=0.92\linewidth]{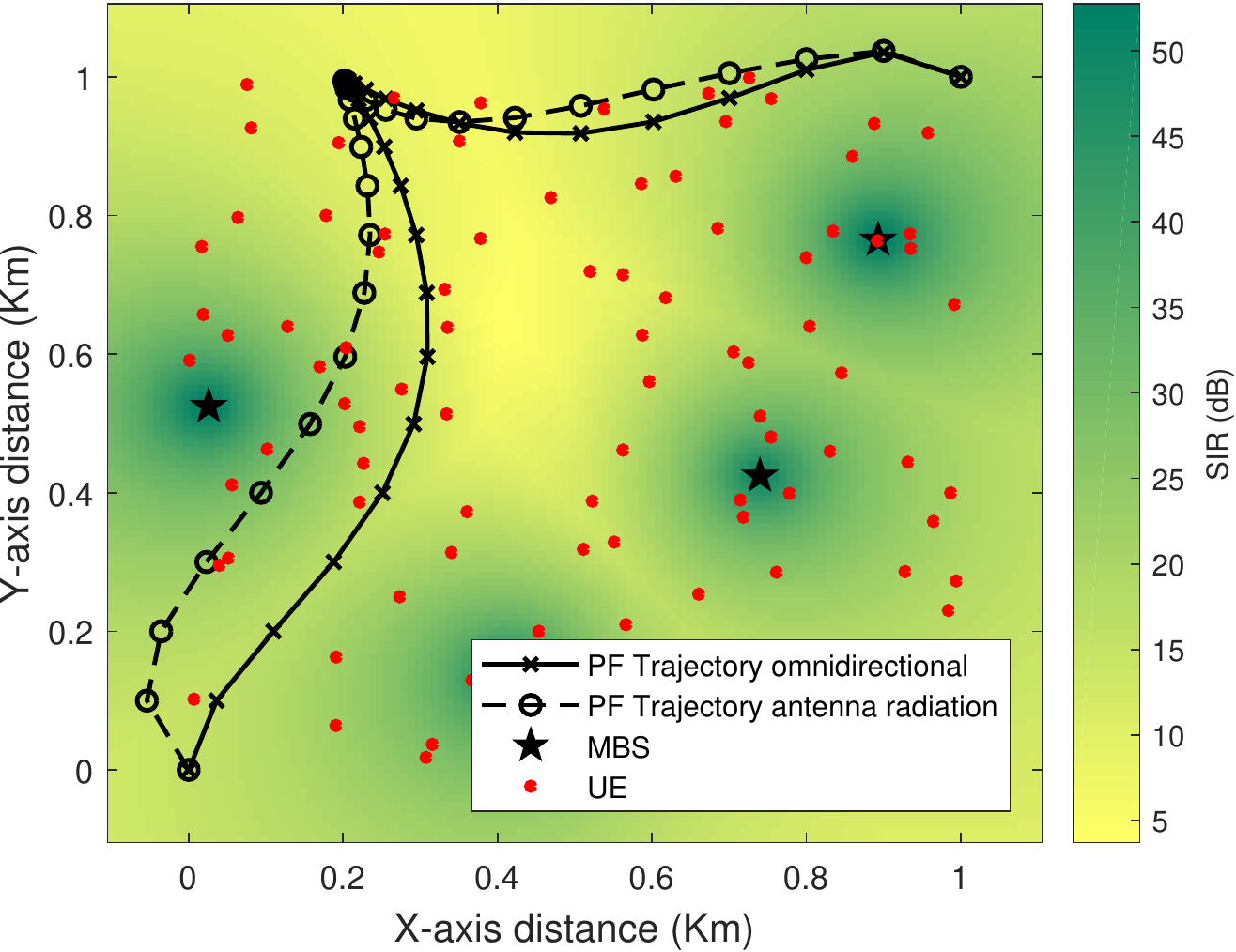}}
			
		\subfloat[]{
			\includegraphics[width=0.92\linewidth]{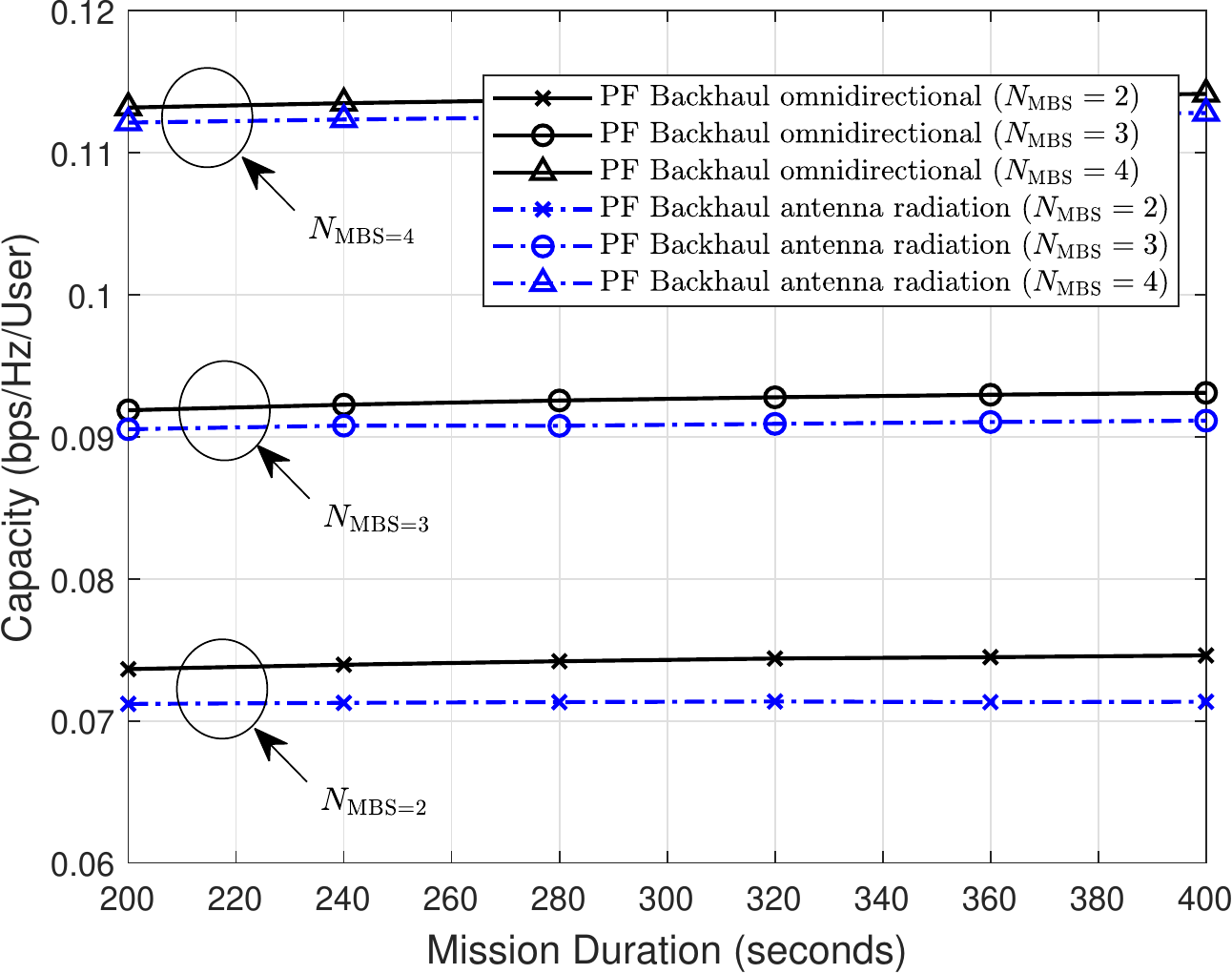}}
          		
		\subfloat[]{
			\includegraphics[width=0.92\linewidth]{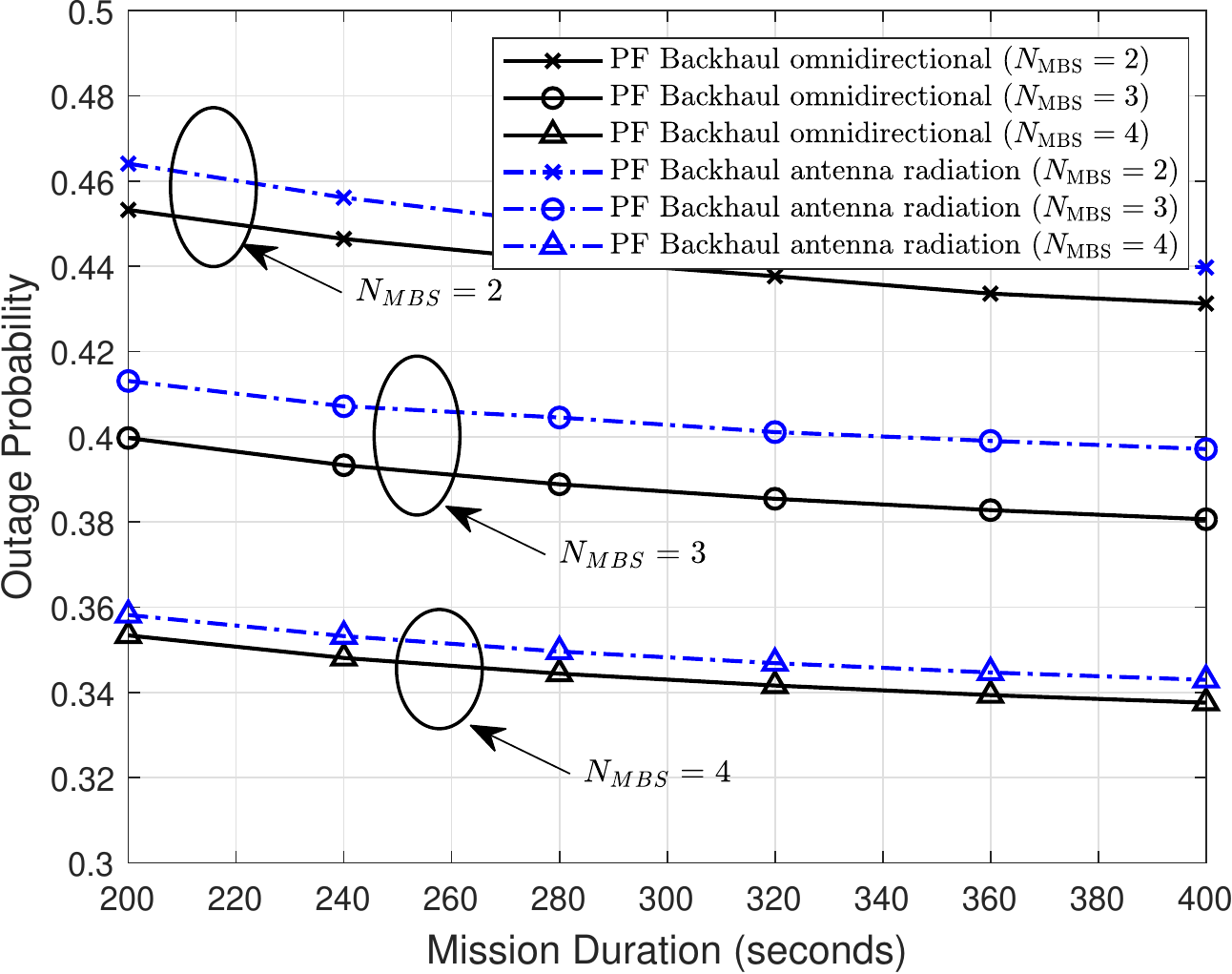}}
		\caption {\textbf{(a) Optimal trajectories of PF rate associated with omnidirectional antenna and considering 3D antenna radiation patterns for \textbf{$T=240$}~s overlapped on SIR (dB) heat map at each discrete point. (b) Network per-user capacity comparison between trajectories with backhaul constraints for omnidirectional and 3D antenna radiation considering OHPLM. 
       (c) Outage Probability comparison between trajectories with backhaul constraints for omnidirectional and 3D antenna radiation considering OHPLM.}. 
        }
		\label{pf_cap_ant2}
	\end{figure}

 In Fig.~\ref{pf_cap_ant2}(a), we plot trajectories of PF rate with and without considering the 3D antenna radiation pattern considering OHPLM. The trajectory generated by taking antenna radiation pattern into consideration follows almost the same path as the other one with omnidirectional antenna scenario. Due to cross dipole antenna orientation, the total gain will be low if the UAV directly hovers over its intended UEs. Hence, it will try to maintain some amount of distance from the associated UEs. Since this phenomenon will introduce degradation in SIRs and hence in capacity, it is understandable that taking antenna radiation into account will result into lower capacity and higher outage performance. Another interesting observation is, the UAV can move closer to MBSs than its omnidirectional counterpart due to the fact that the UEs close to the MBSs also get low throughput due to radiation effect on MBS-UE link.

We also study the capacity and outage performance comparison between backhaul constrained PF trajectories with and without antenna radiation pattern consideration. After observing Fig.~\ref{pf_cap_ant2}(b) and Fig.~\ref{pf_cap_ant2}(c), we can conclude that considering realistic antenna radiation effect into backhaul constraint, introduces lower per UE capacity and higher outage probability. Introducing antenna radiation will provide antenna gains less than 0 dB for some points on the trajectories. Moreover, the UEs close to MBS/UAV will suffer due to radiation/polarization effects, who otherwise would get high throughput because of high SIR in the omnidirectional antenna radiation scenario.     
\section{Concluding Remarks}
In this paper, we study the effects of various different network design issues such as scheduling criteria, UAV mobility constraints,  path loss models, backhaul constraints, and antenna radiation pattern on the trajectory design problem in interference prevalent downlink cellular networks. We first formulate the trajectory optimization problems for different scheduling criteria and solve them using the dynamic programming technique. We also explore and study the capacity and outage probability of the optimal trajectories associated with different network design aspects. We generate smooth trajectories using Bezier curves and show the performance invariability of the smoothed trajectories in terms of cellular user capacity. We explore the performance gain comparison between three path loss models and show that FSPL provides better capacity and outage performance than those of other two path loss models. We compare the network performances of introducing backhaul constraint on the trajectories with PF rate trajectory without constraints. Finally, we explore the effect of introducing 3D antenna radiation pattern on backhaul constraint trajectories assuming circular polarization among the antennas of UAV and MBSs. Our simulation results demonstrate that considering backhaul constraint degrades the network performance. Among the three considered path loss models, OHPLM provides the best network performance while considering the backhaul constraint. Moreover, the antenna polarization pattern results into lower capacity and coverage performance.

We believe our comprehensive study will provide insights into integrating UAVs into cellular networks. As interference limits the overall performance in cellular networks, in future we plan to integrate the effects of interference coordination into trajectory planning.

\acknowledgments
This research was supported by NSF under the grant CNS-1453678.

\bibliographystyle{IEEEtran} 
\bibliography{ref}

\begin{thebibliography}{10}
\providecommand{\url}[1]{#1}
\csname url@samestyle\endcsname
\providecommand{\newblock}{\relax}
\providecommand{\bibinfo}[2]{#2}
\providecommand{\BIBentrySTDinterwordspacing}{\spaceskip=0pt\relax}
\providecommand{\BIBentryALTinterwordstretchfactor}{4}
\providecommand{\BIBentryALTinterwordspacing}{\spaceskip=\fontdimen2\font plus
\BIBentryALTinterwordstretchfactor\fontdimen3\font minus
  \fontdimen4\font\relax}
\providecommand{\BIBforeignlanguage}[2]{{%
\expandafter\ifx\csname l@#1\endcsname\relax
\typeout{** WARNING: IEEEtran.bst: No hyphenation pattern has been}%
\typeout{** loaded for the language `#1'. Using the pattern for}%
\typeout{** the default language instead.}%
\else
\language=\csname l@#1\endcsname
\fi
#2}}
\providecommand{\BIBdecl}{\relax}
\BIBdecl

\bibitem{Bulut}
E.~Bulut and I.~G{\"{u}}ven{\c{c}}, ``Trajectory optimization for
  cellular-connected {UAVs} with disconnectivity constraint,'' in \emph{Proc.
  {IEEE} ICC Workshops}, Kansas City, MO, May 2018, pp. 1--6.

\bibitem{dp}
D.~P. Bertsekas, \emph{Dynamic Programming and Optimal Control}, 3rd~ed.\hskip
  1em plus 0.5em minus 0.4em\relax Belmont, MA, USA: Athena Scientific, 2005,
  vol.~I.

\bibitem{adhoc}
J.~Yoon, Y.~Jin, N.~Batsoyol, and H.~Lee, ``Adaptive path planning of uavs for
  delivering delay-sensitive information to ad-hoc nodes,'' in \emph{Proc. IEEE
  Wireless Commun. Netw. Conf. (WCNC)}, Mar. 2017, pp. 1--6.

\bibitem{rajeev}
R.~Gangula, P.~de~Kerret, O.~Esrafilian, and D.~Gesbert, ``Trajectory
  optimization for mobile access point,'' in \emph{Proc. IEEE Asilomar Conf.
  Sig. Syst. Comp.}, Pacific Grove, CA, Oct. 2017, pp. 1412--1416.

\bibitem{gesbert}
R.~{G}angula, D.~{G}esbert, D.-F. {K}{\"u}lzer, and J.~M. {F}ranceschi
  {Q}uintero, ``{A} landing spot approach for enhancing the performance of
  {UAV}-aided wireless networks,'' in \emph{Proc. IEEE Int. Conf. Commun.
  (ICC)}, Kansas City, MO, May 2018.

\bibitem{moin}
M.~M.~U. Chowdhury, E.~Bulut, and I.~G{\"{u}}ven{\c{c}}, ``Trajectory
  optimization in \textsc{UAV-A}ssisted cellular networks under mission
  duration constraint,'' in \emph{Proc. {IEEE} Radio Wireless Symposium (RWS)},
  Orlando, FL, Jan. 2019.

\bibitem{rui1}
Q.~Wu, Y.~Zeng, and R.~Zhang, ``Joint trajectory and communication design for
  multi-uav enabled wireless networks,'' \emph{IEEE Trans. Wireless Commun.},
  vol.~17, no.~3, pp. 2109--2121, Mar. 2018.

\bibitem{rui2}
Y.~Zeng, X.~Xu, and R.~Zhang, ``Trajectory design for completion time
  minimization in uav-enabled multicasting,'' \emph{IEEE Transactions on
  Wireless Communications}, vol.~17, no.~4, pp. 2233--2246, Apr. 2018.

\bibitem{rui3}
Y.~Zeng and R.~Zhang, ``Energy-efficient uav communication with trajectory
  optimization,'' \emph{IEEE Transactions on Wireless Communications}, vol.~16,
  no.~6, pp. 3747--3760, June 2017.

\bibitem{saad}
U.~Challita, W.~Saad, and C.~Bettstetter, ``Deep reinforcement learning for
  interference-aware path planning of cellular connected {UAVs},'' in
  \emph{Proc. IEEE Int. Conf. Commun. (ICC)}, Kansas City, MO, May 2018.

\bibitem{jiang}
F.~Jiang and A.~L. Swindlehurst, ``Optimization of {UAV} heading for the
  ground-to-air uplink,'' \emph{IEEE J. Sel. Areas Commun. (JSAC)}, vol.~30,
  no.~5, pp. 993--1005, June 2012.

\bibitem{merwaday}
A.~Merwaday, A.~Tuncer, A.~Kumbhar, and I.~Guvenc, ``Improved throughput
  coverage in natural disasters: Unmanned aerial base stations for
  public-safety communications,'' \emph{IEEE Veh. Technol. Mag.}, vol.~11,
  no.~4, pp. 53--60, 2016.

\bibitem{bezier1}
\BIBentryALTinterwordspacing
O.~K. Sahingoz, ``Generation of bezier curve-based flyable trajectories for
  multi-uav systems with parallel genetic algorithm,'' \emph{Journ. of
  Intelligent {\&} Robotic Syst.}, vol.~74, no.~1, pp. 499--511, Apr 2014.
  [Online]. Available: \url{https://doi.org/10.1007/s10846-013-9968-6}
\BIBentrySTDinterwordspacing

\bibitem{bezier2}
R.~Winkel, ``Generalized bernstein polynomials and bézier curves: An
  application of umbral calculus to computer aided geometric design,''
  \emph{Advances in Applied Mathematics}, vol.~27, no.~1, pp. 51 -- 81, 2001.

\bibitem{hata}
Y.~Singh, ``Comparison of {Okumura, Hata and COST-231} models on the basis of
  path loss and signal strength,'' \emph{Int. J. Comp. Appl.}, vol.~59, no.~11,
  pp. 37--41, Dec. 2012.

\bibitem{mixturepl}
M.~M. Azari, F.~Rosas, A.~Chiumento, and S.~Pollin, ``Coexistence of
  terrestrial and aerial users in cellular networks,'' in \emph{Proc. IEEE
  Globecom Workshops (GC Wkshps)}, Singapore, Dec 2017, pp. 1--6.

\bibitem{fspl}
A.~Al-Hourani, S.~Kandeepan, and A.~Jamalipour, ``Modeling air-to-ground path
  loss for low altitude platforms in urban environments,'' in \emph{Proc. IEEE
  Globecom}, Austin, TX, Dec 2014, pp. 2898--2904.

\bibitem{backhaul}
H.~C. Nguyen, R.~Amorim, J.~Wigard, I.~Z. Kovács, T.~B. Sørensen, and P.~E.
  Mogensen, ``How to ensure reliable connectivity for aerial vehicles over
  cellular networks,'' \emph{IEEE Access}, vol.~6, pp. 12\,304--12\,317, 2018.

\bibitem{chetan}
K.~A. Chethan and C.~R. Murthy, ``An iterative re-weighted minimization
  framework for resource allocation in the single-cell relay-enhanced
  \textsc{OFDMA} network,'' in \emph{Proc. IEEE Int. Workshop on Signal
  Processing Adv. Wireless Commun. (SPAWC)}, Edinburgh, UK, July 2016, pp.
  1--6.

\bibitem{sir}
V.~A. Aalo, G.~P. Efthymoglou, T.~Soithong, M.~Alwakeel, and S.~Alwakeel,
  ``Performance analysis of multi-hop amplify-and-forward relaying systems in
  rayleigh fading channels with a poisson interference field,'' \emph{IEEE
  Trans. Wireless Commun.}, vol.~13, no.~1, pp. 24--35, Jan. 2014.

\bibitem{3gpp}
\BIBentryALTinterwordspacing
3GPP, Technical Specification (TS) 36.777, 2018. [Online]. Available:
  \url{https://portal.3gpp.org/desktopmodules/Specifications/\\SpecificationDetails.aspx?specificationId=3231}
\BIBentrySTDinterwordspacing

\bibitem{priyanka_3d}
J.~{Chen}, D.~{Raye}, W.~{Khawaja}, P.~{Sinha}, and I.~{Guvenc}, ``Impact of
  3\textsc{D} \textsc{UWB} antenna radiation pattern on air-to-ground drone
  connectivity,,'' in \emph{Proc. IEEE Vehc. Technol. Conf. (VTC).}, Chicago,
  IL, Sep 2018.

\bibitem{mmimo}
P.~Chandhar, D.~Danev, and E.~G. Larsson, ``Massive \textsc{MIMO} for
  communications with drone swarms,'' \emph{IEEE Trans. Wireless Commun.},
  vol.~17, no.~3, pp. 1604--1629, Mar. 2018.

\bibitem{Balanis}
C.~A. Balanis, \emph{Antenna Theory: Analysis and Design}.\hskip 1em plus 0.5em
  minus 0.4em\relax New York, NY, USA: Wiley-Interscience, 2005.

\end{thebibliography}



\end{document}